\documentclass[10pt, conference, compsocconf]{IEEEtran}
\IEEEoverridecommandlockouts

\usepackage{makeidx}  
\usepackage{graphicx}
\usepackage{acronym}
\usepackage{subfigure}
\usepackage{booktabs}
\usepackage{algorithm}
\usepackage{algpseudocode}
\usepackage{url}
\usepackage{amstext}
\usepackage{amsmath}

\acrodef{P2P}{Peer-to-Peer}
\acrodef{TTL}{Time-To-Live}
\acrodef{ECC}{Edge Clustering Coefficient}

\begin{document}
 
%
%
\title{On the Topology Maintenance of Dynamic P2P Overlays through Self-Healing Local Interactions}
\author{\IEEEauthorblockN{Stefano Ferretti}
\IEEEauthorblockA{Department of Computer Science and Engineering, University of Bologna\\
Mura Anteo Zamboni 7, Bologna, Italy\\
s.ferretti@unibo.it}
}

\maketitle              

\begin{abstract}
This paper deals with the use of self-organizing protocols to improve the reliability of dynamic Peer-to-Peer (P2P) overlay networks. We present two approaches, that employ local knowledge of the $2$nd neighborhood of nodes. The first scheme is a simple protocol requiring interactions among nodes and their direct neighbors. The second scheme extends this approach by resorting to the Edge Clustering Coefficient (ECC), a local measure that allows to identify those edges that connect different clusters in an overlay. A simulation assessment is presented, which evaluates these protocols over uniform networks, clustered networks and scale-free networks. Different failure modes are considered. Results demonstrate the viability of the proposal.
\end{abstract}
%

\section{Introduction}

The organization of \ac{P2P}~networks as unstructured overlays has been recognized as an interesting solution in several domains. According to it, links among peers are established arbitrarily, not depending on the typology of nodes, on some characteristics they possess, or on the information they maintain \cite{Luciano1,EberspacherS05a,simplex,Leitao}.
The ease in building and maintaining unstructured overlays make these solutions particularly appealing when the P2P system is very dynamic \cite{gridpeer,shimada}. 
One of the most fascinating aspects of these distributed approaches is that peers can execute local strategies in order to maintain some global properties of the overall network through decentralized interactions. These global properties are usually referred as self-* properties (e.g., self-organization, self-adaptation, self-management). Among them, self-healing figures as a key characteristics to improve the dependability of the managed infrastructure.
Self-healing is not novel in networks. It is an interesting approach to cope with the general problem of providing network resilience \cite{doerr}.
It has been a long time since self-healing ring topologies have been introduced.
In the domain of P2P (and networks), several works concerned with this issue have been proposed \cite{chaudhry,simplex,Pournarasb}.

However, in P2P systems, certain network properties are guaranteed usually on the steady state. Thus, 
it may happen that they disappear in case of multiple node departures \cite{Leonard:2005}. For instance, the overlay might get partitioned upon failure of links connecting different clusters. 
Alternatively, some important links might be lost that were playing a main role to keep a low network diameter; for instance, in small worlds there are links among distant nodes, that strongly reduce the average shortest path length.
Although the P2P network is unstructured, it has a topology that provides certain characteristics. These should be maintained, at least up to a certain extent, in order to provide some guarantees and the ability of the network to spread contents.

In this work, we study the effectiveness of a decentralized self-healing algorithm to provide resilience of the P2P overlay. 
Te idea is to track the list of $2$nd neighbors, i.e.~the neighbors of the neighbors of a node. Upon failure of a neighbor, each node reacts trying to keep itself connected to the rest of the network.
We consider two variants of this approach. 

According to the first protocol (hereinafter referred as $P_{2n}$), 
each node $n$ checks if some $2$nd neighbor is no more reachable. This possible situation triggers a decentralized procedure to guarantee that all those connections with $2$nd neighbors, which are lost due to the removal of the intermediate node, are get back.
It suffices that only one node, in a set of connected neighbors, creates a novel link with each lost $2$nd neighbor \cite{simplex13}.
The interesting aspect of this mechanism is that it does not require full knowledge of the overlay network \cite{massoulie,VoulgarisGS05}. 

The second protocol (hereinafter referred as $P_{ECC}$) is similar to the one above, i.e., upon a node failure, its neighbors try to cope with possible network partitions by creating novel links to their lost $2$nd neighbors. However, instead of executing a randomized distributed link creation procedure, each node exploits a measure it has of the \ac{ECC} of the lost link.
\ac{ECC} is a measure which has been proposed in \cite{radicchi2004}. It counts the number of triangles to which a given edge belongs, divided by the number of triangles that might potentially include it, given the degrees of the adjacent nodes.
ECC is thus a local measure, and it allows to identify those edges that connect different clusters in an overlay. 
Indeed, the lower the ECC of an edge the more likely two nodes at the ends of the link would become disconnected if that edge fails (i.e.~there are few other short paths that connect them).
In fact, edges connecting nodes in different communities are included in few or no triangles, and tend to have small values of ECC. On the other hand many triangles exist within clusters. Hence, ECC is a measure of how inter-communitarian a link is.

According to $P_{ECC}$, upon a link removal at a node $n$, due to the failure of one of its neighbors, $n$ checks the ECC of this link. In particular, $n$ decides to activate the failure recovery procedure with a probability which is inversely proportional to this ECC value, i.e.~the more the link was part of triangles, the lower such a probability. 
The recovery procedure consists in creating links with the lost $2$nd neighbors.
With respect to $P_{2n}$, a node might avoid to activate the self-healing procedure for those lost links with higher ECC values.
Not only, with the aims of preserving the network topology and of limiting the potential growth on the number of links in the network, a link removal phase is included in the protocol. Basically, it removes (with a certain probability) links with higher ECC values, associated to nodes with a degree exceeding their target degree.

A simulation assessment is presented that studies the protocols over uniform networks, where links are created by randomly choosing nodes as neighbors, clustered networks and scale-free networks. 
Results demonstrate that the presented self-healing approaches preserve networks connectivity, despite node churn and targeted attacks.

%

The remainder of this paper is organized as follows. Section \ref{sec:prot} presents the P2P protocol. Section \ref{sec:eval} describes the simulation environment and discusses the obtained results. Finally, Section \ref{sec:conc} provides some concluding remarks.

\section{Self-Healing Protocols}\label{sec:prot}

We already mentioned that each node $n$ has a certain degree, i.e.~the amount of $1$st neighbors. 
The list of these $n$'s $1$st neighbors is denoted with $\Pi_n$, while the degree of $n$ is denoted with $|\Pi_n|$. $n$ also maintains the list of its 2nd neigh\-bors, $\Pi^2_n$, i.e.~nodes distant $2$ hops from $n$. 
Every time the list of 1st neighbors $\Pi_n$
changes, due to some node arrival or departure, $n$ informs its other 1st neighbors of this update.
With $\Pi^2_{n|m} = \Pi_m - \Pi_n$, we identify the $n$'s 2nd neighbors which can be reached through $m$. Hence, $\Pi^2_n = \cup_{k \in \Pi_n} \Pi^2_{n|k}$.

\subsection{Protocol $P_{2n}$: Use of the $2$-Neighborhood}

According to this self-healing protocol, upon a neighbor departure, each node $n$ is able to understand if some 2nd neighbor is no more reachable. 
If this is the case, $n$ creates a link with it.
Algorithms \ref{alg:active}--\ref{alg:passive} sketch the related pseudo-code.
In particular, when a node $f \in \Pi_n$ fails, $\forall p \in \Pi_f$ there are three possible cases.\\
1) $p \in \Pi_n$ : $n$ and $p$ are neighbors. In this case there is nothing to do (at $n$).\\
2) $p \notin \Pi_n$, but $p \in \Pi^2_n$ since $p \in \Pi^2_{n|q}$ for some $q \in \Pi_n, q \neq f$: $p$ is still a 2nd neighbor of $n$; also in this case there is nothing to do.\\
3) $p \notin \Pi_n, p \notin \Pi^2_n$ : $p$ is no more a $1$st or 2nd neighbor of $n$. In this case, $n$ takes part to the distributed procedure to create a link with $p$ (see Algorithm \ref{alg:active}).

In essence, links are created among nodes in different clusters which were connected through $f$ only.
Each node $n$ keeps a threshold value for its degree, to avoid that its degree grows out of control. (However, this threshold must not be too low, otherwise it would contrast the creation of additional links, and this might generate network partitions.)
Moreover, in order to diminish the probability that multiple nodes of the same cluster attempt to create a novel link with the same node $p$ at the same time, a classic contention-based approach is used, so that each node $n$ waits for a random time before transmitting messages (Algorithm \ref{alg:active}, line \ref{code:wait}). 
Then, upon reception of a message from a node $p$ asking $n$ to become neighbors, $n$ accepts the request only if $p$ is not a 1st or 2nd neighbor of $n$ (it is possible that some of its neighbors just created a link with $p$; see Algorithm \ref{alg:passive}).

\begin{algorithm}[htbp]
\caption{$P_{2n}$: Active behavior at $n$ upon failure of $f$}
\label{alg:active}
\begin{algorithmic}[1]
\State $P \gets \{ p \in \Pi_f | \ p \notin \Pi_n, p \notin \Pi^2_n \}$
\Statex
\While {($P \neq \emptyset) \wedge (| \Pi_n | \leq \text{thresholdDegree})$}\label{code:control}
  \State wait random time \label{code:wait}
  \State $p \gets$ extract random node from $P$ \label{code:extract}
  \State send link creation request to $p$\label{code:req}
\EndWhile
\end{algorithmic}
\end{algorithm}
\begin{algorithm}
\caption{$P_{2n}$: Passive behavior at $n$}
\label{alg:passive}
\begin{algorithmic}[1]
\Require message from $p$ answering a link creation request
  \If {answer is OK}\label{code:ans_b}
    \State sendAll($\Pi_n$, ``novel link $(n, p)$'')
    \State add $p$ to $\Pi_n$
  \EndIf\label{code:ans_e}
\Statex 
\Require message from $q \in \Pi_n$: novel link $(q, m), m \in P$
  \State extract $m$ from $P$\label{code:l_creation}
\Statex 
\Require message from $p$ with a link creation request
  \If{$p \in \Pi^2_n$}\label{code:l_creat_req_b}
    \State send refuse message
  \Else
    \State send accept message
    \State sendAll($\Pi_n$, ``novel link $(n, p)$'')
    \State add $p$ to $\Pi_n$
  \EndIf\label{code:l_creat_req_e}
\end{algorithmic}
\end{algorithm}

\subsection{Protocol $P_{ECC}$: Edge Clustering Coefficient}

This protocol is an extension of $P_{2n}$, and it is based on the idea of exploiting the importance of failed links, so as to identify those that, once failed, must be replaced with novel ones. 
In complex network theory, several centrality measures have been introduced to characterize the importance of a node or a link in a network, e.g.~betweenness centrality, or to detect different communities and identify their boundaries in the net \cite{Bader:2007,girvan,Goncalves:2012,Newman200539}.
The calculation of these metrics usually involves a full (or partially full) knowledge about the whole network. 
Conversely, the aim of this work is to preserve connectivity without such a global knowledge \cite{simplex13,massoulie,VoulgarisGS05}.

The Edge Clustering Coefficient (ECC) has been defined in analogy with the usual node clustering coefficient, but it is referred to an edge of the network \cite{radicchi2004}. It measures the number of triangles
to which a given edge belongs, divided by the number of triangles that might potentially include it, given the degrees of the adjacent nodes. More formally, given a link $(n, m)$ connecting node $n$ with node $m$, the edge clustering coefficient $ECC_{n,m}$ is
$$ECC_{n,m} = \frac{T_{n,m}}{min((|\Pi_n| - 1), (|\Pi_m| - 1))},$$
where $T_{n,m}$ is the number of triangles built on that edge $(n,m)$, and $min((|\Pi_n | -1), (|\Pi_m | -1))$ is the amount of triangles that might potentially include it. We add the constraint that this measure is $0$ when there are no possible triangles at one of the nodes, i.e.~when $min((|\Pi_n | - 1), (|\Pi_m | - 1))=0$.

The idea behind the use of this quantity is that edges connecting nodes in different communities are included in few or no triangles, and tend to have small values of $ECC_{n,m}$. On the other hand many triangles exist within clusters. Hence the coefficient $ECC_{n,m}$ is a measure of how inter-communitarian a link is.

Thus, based on this notion of $ECC_{n,m}$, the protocol $P_{ECC}$ works as follows. (Algorithm \ref{alg:ecc_active} shows the pseudo-code of the active behavior, since the passive behavior is similar to Algorithm \ref{alg:passive}).
Each node $n$ knows its $2$nd neighbors, i.e.~$1$st neighbors of its neighbors; thus, it can understand if some triangle exists that includes itself. 
Indeed, let say that three nodes $n, m, p$ create a triangle. Then, $n$ has $m,p$ in its neighbor list $\Pi_n$ (and the same happens for the two other nodes). When $n$ sends its list $\Pi_n$ to $m$ and $p$, they recognize that there is a common neighbor (i.e.~$n$) that creates a triangle.
If one of the three nodes would fail in the future, the other two nodes will understand automatically that the triangle no longer exists.

\begin{algorithm}[thbp]
\caption{$P_{ECC}$: Active behavior at $n$ upon failure of $f$}
\label{alg:ecc_active}
\begin{algorithmic}[1]
\State $P \gets \{ p \in \Pi_f | \ p \notin \Pi_n, p \notin \Pi^2_n \}$
\Statex
\If{random() $> ECC_{n,f}$}\label{code:ecc}
 \While {($P \neq \emptyset) \wedge (| \Pi_n | \leq \text{thresholdDegree})$}
  \State wait random time 
  \State $p \gets$ extract random node from $P$ 
    \State send link creation request to $p$
 \EndWhile
\EndIf
\Statex 
\Require $(|\Pi_n| \gg |\Pi_n|_{target})\ \wedge \ (L_{n} \gg L_{n, target})$
\State Remove at most $r$ links with $ECC > T_{ECC}$
\end{algorithmic}
\end{algorithm}

When a node, say $f$ fails, each neighbor $n \in \Pi_f$ checks the value $ECC_{n,f}$.
Depending on this value, a reconfiguration phase may be executed. The idea is that the higher the ECC the lower the need to create novel links to keep the network connected, since that link was part of multiple triangles. This decision is taken probabilistically, i.e.~the lower $ECC_{n,f}$ the more probable that the rest of the procedure is executed (line \ref{code:ecc}, Algorithm \ref{alg:ecc_active}).

If this is the case, $n$ checks if its $2$nd neighbors ($\Pi^2_{n|f}$), reached formerly through $f$, still remain in its $2$nd neighborhood; otherwise it creates links with them, as in $P_{2n}$. 

Due to the overlay reconfiguration, it is expected that the degree of a node changes (suddenly, in some cases). Indeed, the goal of the self-healing reconfiguration scheme is that the network should evolve to react to nodes arrivals and departures. For instance, if a hub goes down for some reason, it is likely that its past neighbors would create more links in order to maintain the overlay connected.
Thus, it might happen that the total number of links augments, due to the parallel activity of nodes, and this can alter the network topology. In $P_{ECC}$, this is more probable when there is a low network clustering, with few triangles.  

To overcome this possible problem, a periodical check is accomplished on the growth of links at each node and its neighborhood. 
Thus, periodically each node $n$ checks its actual degree $|\Pi_n|$ and the actual number of links in its neighborhood $L_{n}$, i.e.~the sum of all different links departing from $\Pi_{n} \cup \{n\}$.
These values are compared with two values that $n$ stores, related to the target degree $|\Pi_n|_{target}$ and a target number of links in the $n$'s neighborhood $L_{n, target}$.
By monitoring the amount of links in its neighborhood, $n$ obtains an approximate understanding of how the network is evolving.
(These two values are periodically updated, based on values assumed in a window time interval.)

Once it has been noticed that there has been an important increment on the amount of links in some portion of the network, then the nodes with the higher variations on their degrees check if some links (i.e.~those with higher ECC values) can be removed.
Indeed, if the difference between the target values and the actual ones surpasses a given threshold, then the node $n$ invokes a procedure that removes its $r$ links with higher ECC values (larger than a threshold value $T_{ECC}$), if there are any. (In the simulations, we consider $r=1$ since it suffices to control the rate of the periodical check to increase/decrease the number of links that can be removed.)

\section{Performance Evaluation}\label{sec:eval}

To assess the proposed protocols, we simulated them over different network topologies. In the following, we describe the simulation settings and the obtained results over uniform networks, clustered networks and scale-free networks. 
Actually, during our experimental evaluation we employed also random graphs, obtaining results comparable to those obtained for uniform networks (hence, for the sake of brevity we do not show them in the paper).
The considered approaches are $P_{2n}$, $P_{ECC}$ and ``none'', which represents the (typical) situation when peers do not react to node disconnections, simply assuming that other links will be created upon arrivals of novel nodes.

\subsection{Simulation Details}

The simulator was written in GNU Octave.
We present averages of obtained results from a corpus of $20$ simulations for the same scenario. During the simulations we removed the transient from the analyzed logs. All the configuration parameters were varied; we present here results for some particular configuration settings, since those obtained for different ones were comparable to those we will show.
As mentioned, three types of unstructured networks were considered. 

\subsubsection{Uniform Networks}
Uniform networks are those where all nodes start with the same degree. Then, due to node failures and arrivals (and the reconfiguration imposed by the P2P protocol), the node degree might change.
We varied the initial degree of nodes. 

\subsubsection{Clustered Networks}
These self-healing protocols are thought for those P2P overlays that have important links that connect different parts of the network; thus, it is interesting to observe how the protocol performs over nets composed of different connected clusters.
In these simulations, network clusters were set to be of the same size.
We set two different parameters to create the network. The first parameter is the probability $\gamma$ of creating a link among nodes of the same cluster. Each node is linked to another node of the same cluster with a probability $\gamma$; hence, inside a cluster, nodes are organized as a classic random graph.
As to inter-cluster links, the amount of links created between the two clusters was determined based on a certain probability $\omega$ times the number of nodes in the clusters (i.e.~each node has a probability $\omega$ of having a link with each external cluster).

\subsubsection{Scale-Free Networks}
As to scale-free networks, these are characterized by a degree distribution following a power law. 
The presence of hubs, i.e.~nodes with degrees higher than the average, has an important impact on the connectivity of the net. To build scale-free networks, our simulator implements the construction method proposed in \cite{Aiello00arandom}.

\subsubsection{Simulation of Node Churn and Failures}

Upon a node failure, all its links with other nodes are removed. Then, the node passes to an inactive state; it can be selected further on to simulate a novel node arrival.
Thus, a node arrival is simulated by changing the state of a randomly selected inactive node to pass to the active state. This event triggers the creation of novel links with other randomly selected nodes. 
Different joining procedures were executed, depending on the network topology under investigation. The idea was to adopt a join mechanism that would maintain the topology unaltered.
Thus, as concerns uniform networks, a random set of neighbors was selected, whose size was equal to the fixed degree that characterized the starting network topology.
As to clustered networks, the node was associated to a cluster, and links with nodes in that cluster were randomly created based on the $\gamma$ probability, as in a classic random graph. Then, for each other cluster, the node creates, with probability $\omega$, a link with a random node of that cluster.
Finally, a preferential attachment was utilized for scale-free networks.

Three types of failure schemes were considered. The first mode was based on a random selection of failed nodes, with an amount of failed nodes equal to the amount of joining nodes (we refer to this simulation mode as ``evolution''). This way, the network size remains stable during the simulation. 

A second mode was a ``targeted attack'', meaning that at each step the ``important'' nodes with some specific characteristics were selected to fail. In particular, as concerns uniform and scale-free networks, nodes with the higher degrees were selected to fail. Instead, in clustered networks the selected nodes were those with higher number of links connecting different clusters (the rationale was to augment the probability of disconnecting the clusters).
In this scheme, as in the previous simulation mode the amount of failed nodes per simulation time interval was kept equal to the amount of joining nodes.

As to the third simulation mode, only failures occurred. Thus, each network started with all nodes active, which were (randomly) forced to fail until no active nodes remain in the network. This allows to understand if the self-healing protocols are able to react to situations with high failure rates. We refer to this simulation mode as ``failures only''.

\subsection{Results}

\subsubsection{Uniform Networks}
\begin{figure*}[t]
   \centering
   \includegraphics[width=.48\linewidth]{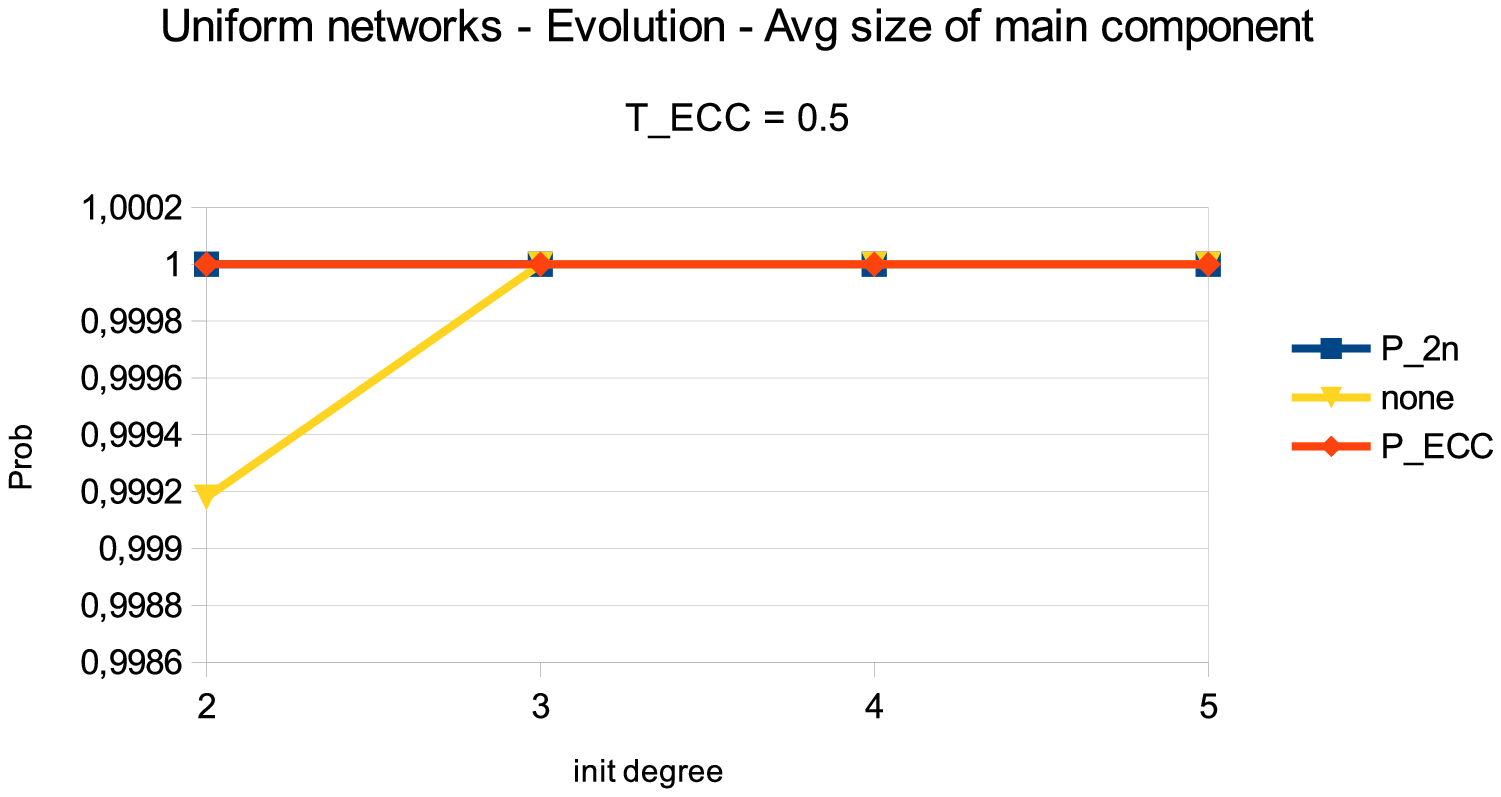}
   \includegraphics[width=.48\linewidth]{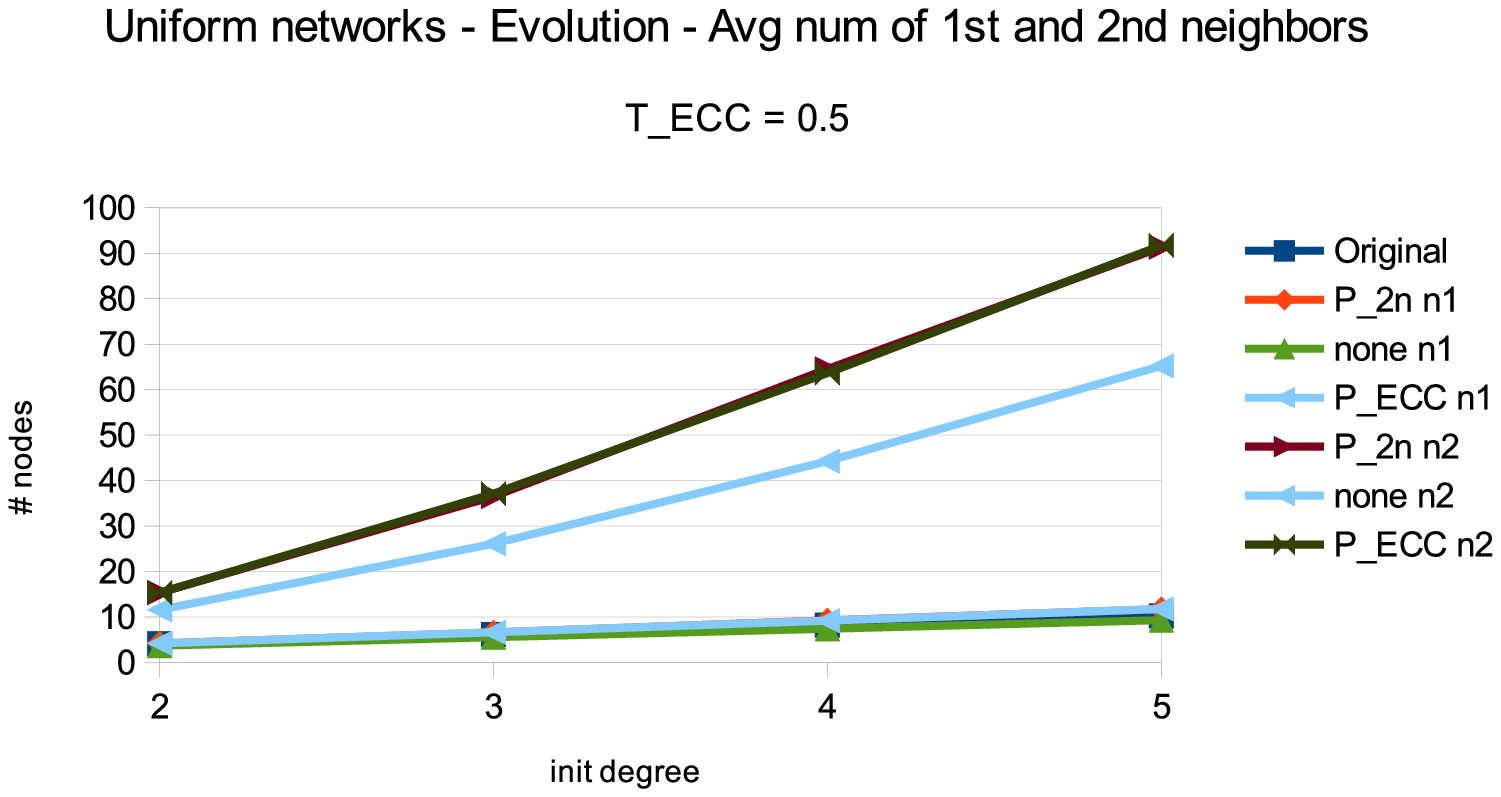}
   \caption{Uniform networks: average size of the main components, average amount of $1$st neighbors (referred as ``n1'') and $2$nd neighbors (referred as ``n2''), during the evolution of the network.}
   \label{fig:unif_evol}
\end{figure*}

Figure \ref{fig:unif_evol} shows results for uniform networks, when the simulation was run with the evolution simulation mode. The chart on the left reports the average size of the main component for the three considered management protocols, while the chart on the right reports the average amount of $1$st and $2$nd neighbors. In this case, as expected the failure of nodes does not create particular problems, since others arrive in the meantime. Thus, the topology remains pretty much unvaried. It is interesting to observe that however, when the amount of links is low, a small portion of nodes of the network can remain outside the main component when no failure management mechanisms are employed (see ``none'' curve on the left chart). 
Another interesting aspect is that, while small variations on the average amount of $1$st neighbors is noticed for the three schemes (the ``none'' protocol has a slight lower average value than the other two approaches), the average amount $2$nd neighbors is significantly lower for the ``none'' protocol w.r.t.~$P_{2n}$, $P_{ECC}$. This means that these two self-healing protocols guarantee a higher connectivity for uniform networks.

\begin{figure*}[t]
   \centering
   \includegraphics[width=.48\linewidth]{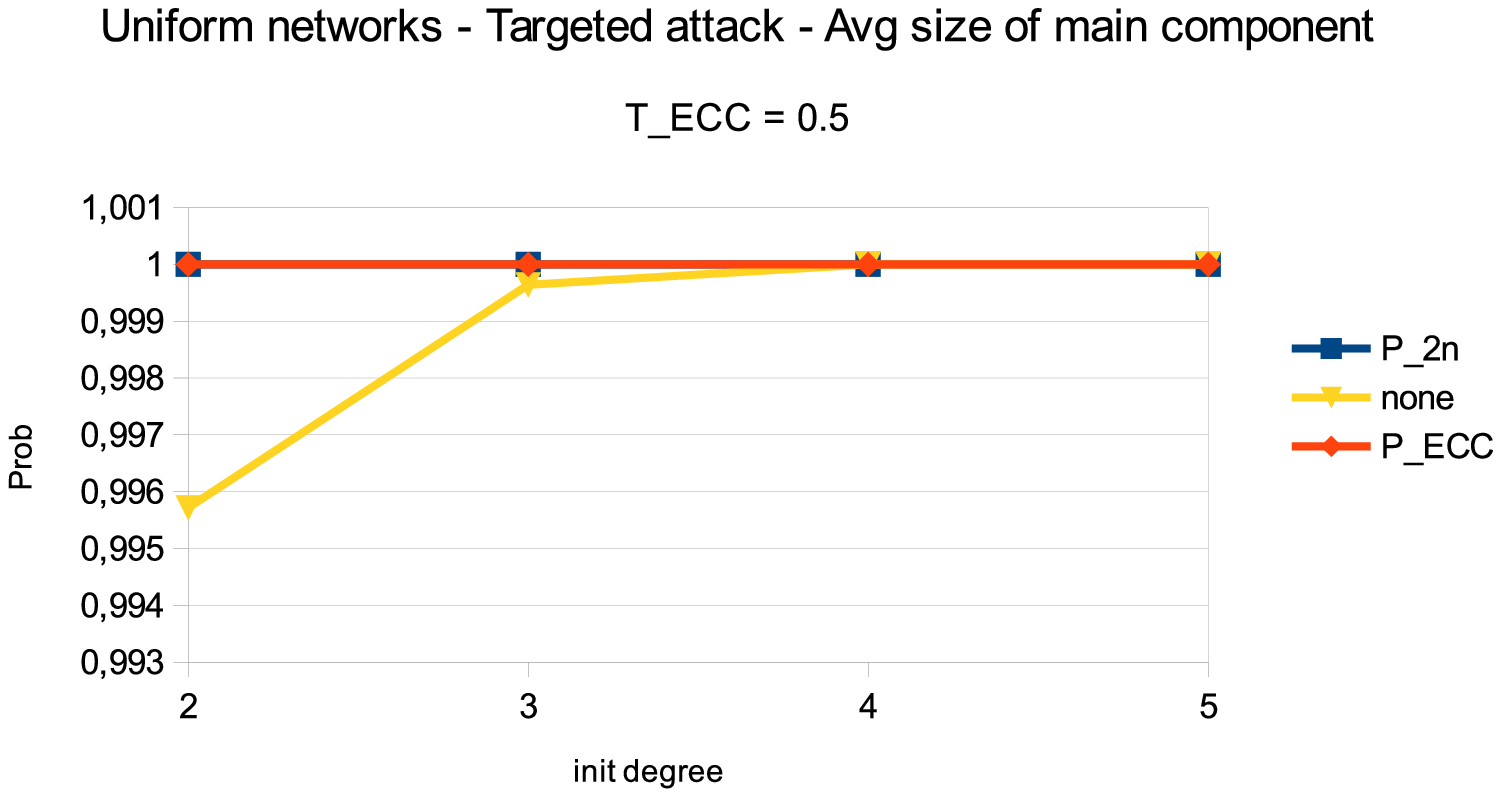}
   \includegraphics[width=.48\linewidth]{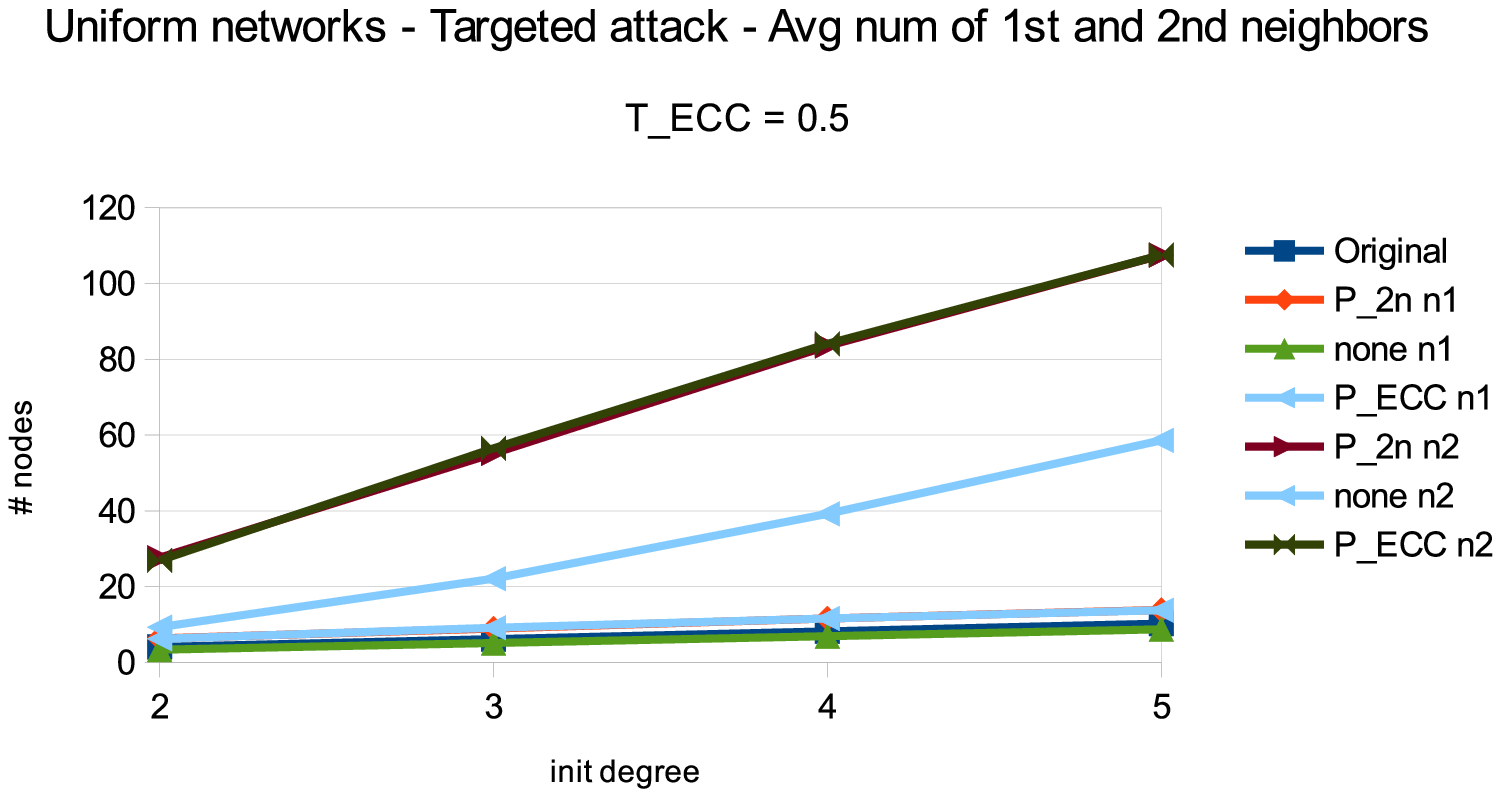}
   \caption{Uniform networks: average size of the main components, average amount of $1$st neighbors (referred as ``n1'') and $2$nd neighbors (referred as ``n2''), under a targeted attack.}
   \label{fig:unif_targ}
\end{figure*}

Similar outcomes are noticed when the targeted attack simulation mode is run, as shown in Figure \ref{fig:unif_targ}.
In fact, in uniform networks there are no important differences between nodes (i.e.~all nodes start with the same degree), thus the selection of the one with highest degree has not a significant impact on the topology. Nevertheless, it is possible to appreciate that the number of $2$nd neighbors decreases for the ``none'' protocol, also w.r.t.~results obtained for the evolution simulation modes. Similarly, in the ``none'' protocol the average size of the main component results lower w.r.t.~that reported in Figure \ref{fig:unif_evol} obtained in the evolution simulations. Conversely, results remain unchanged for $P_{2n}$ and $P_{ECC}$.

\begin{figure*}[t]
   \centering
   \subfigure{\includegraphics[width=.37\linewidth]{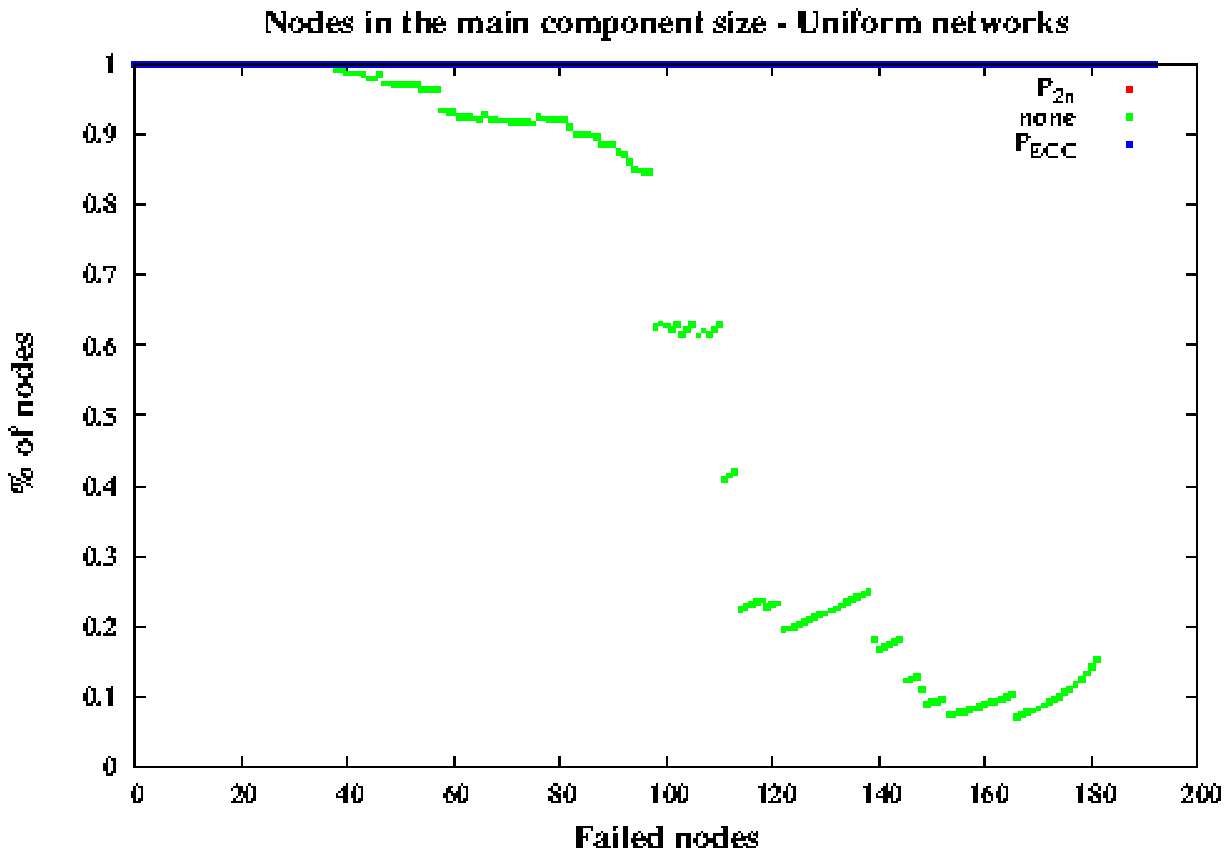}}
      \hspace{1.5cm}
      \subfigure{\includegraphics[width=.37\linewidth]{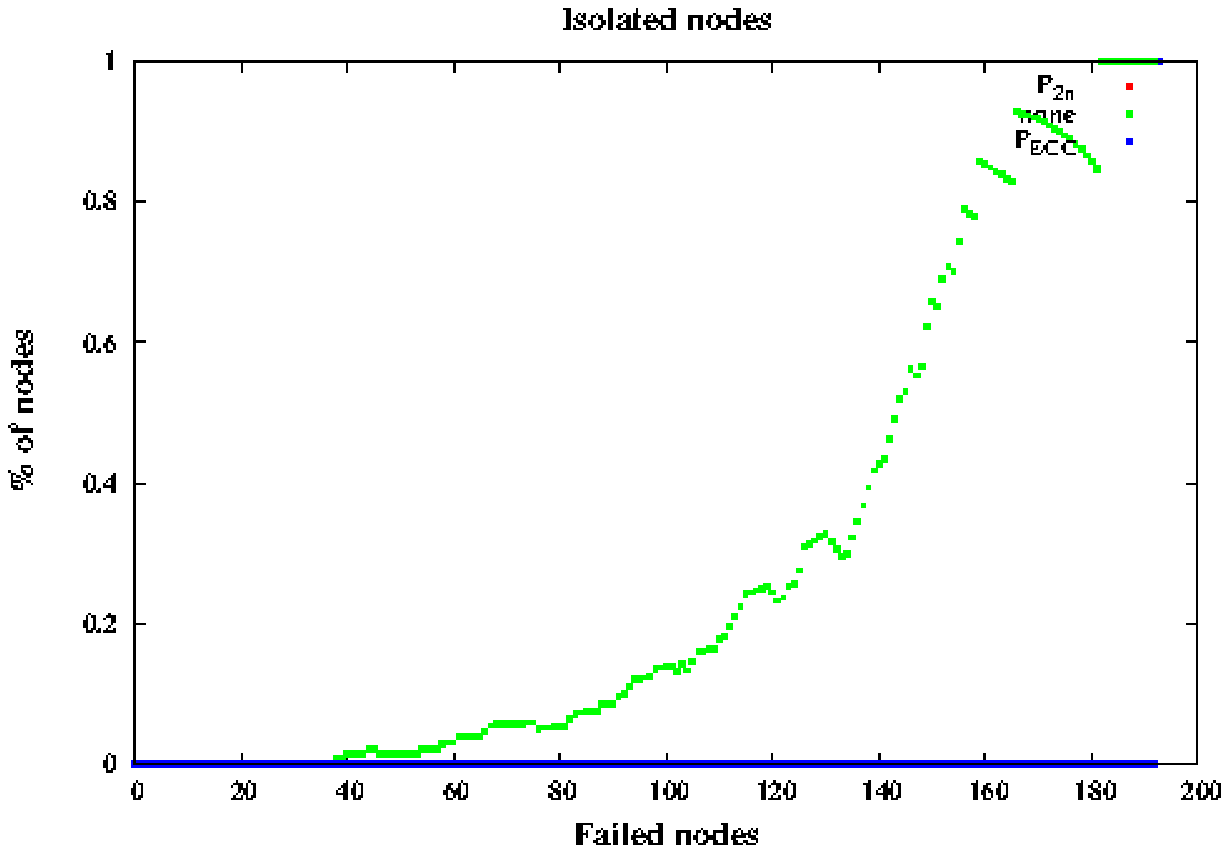}}
   \caption{Uniform networks -- progressive node failures: Amount of nodes in the main component, isolated nodes.}
   \label{fig:unif_fail}
\end{figure*}

Figure \ref{fig:unif_fail} shows results obtained under the ``failures only'' simulations. In particular, the chart on the left shows the amount of nodes that remain in the main component, while nodes continuously fail, for a typical simulation over a uniform network. (We repeated the same experiment multiple times, varying the network size, the initial nodes' degree, and the seed for random generations, obtaining comparable results.)
It is possible to see that, in the ``none'' protocol, at a certain point of the simulation the network gets disconnected and the percentage of active nodes in the main components decreases. Instead, in $P_{2n}$ and $P_{ECC}$, active nodes remain connected in the same, single component. 
This is confirmed by looking at the chart on the right in the same figure, which shows the amount of isolated nodes. While the percentage of isolated nodes increases in the ``none'' scheme, no nodes remain isolated for the other two protocols.

\subsubsection{Clustered Networks}
\begin{figure*}[t]
   \centering
   \includegraphics[width=.48\linewidth]{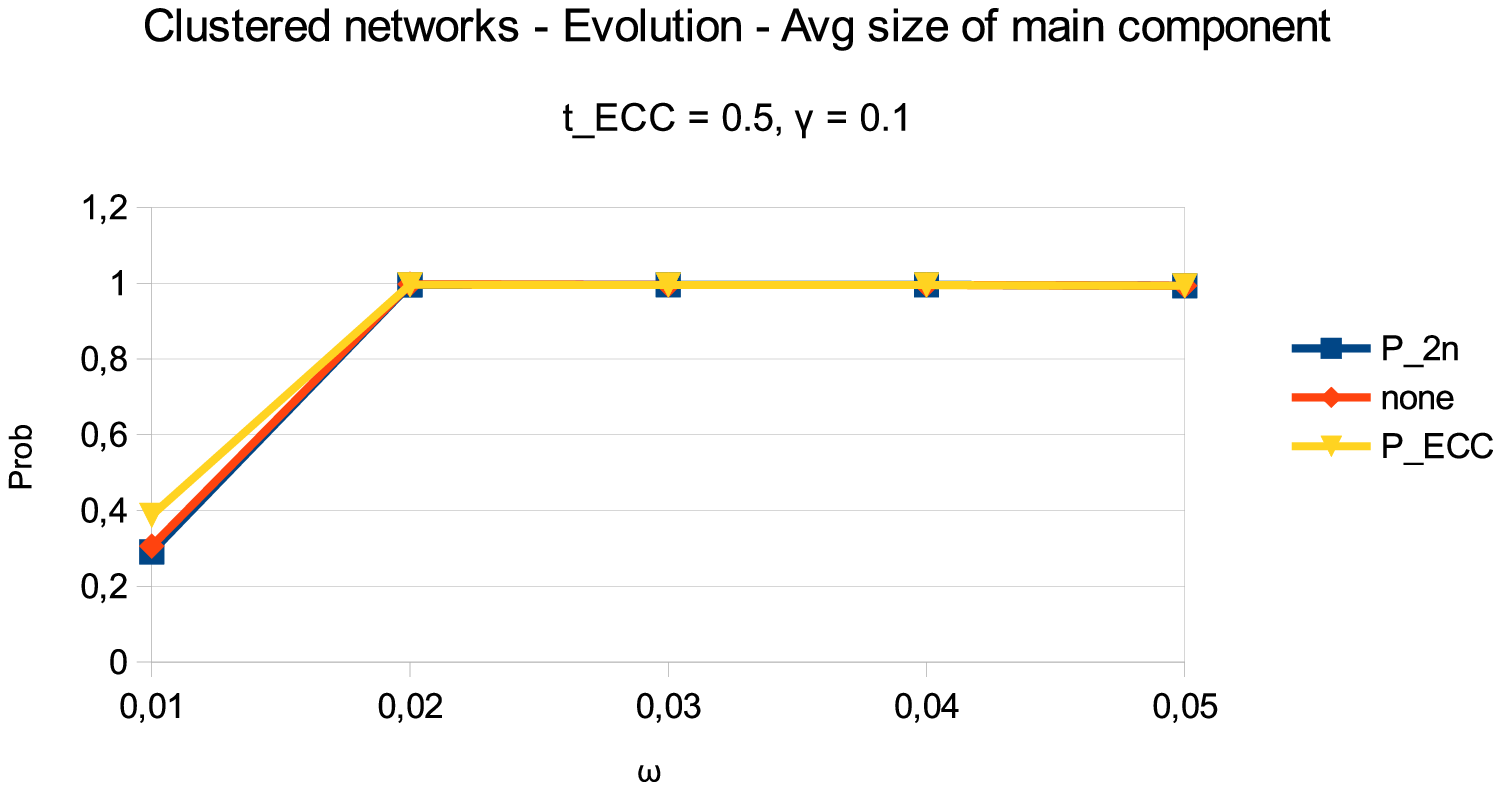}
   \includegraphics[width=.48\linewidth]{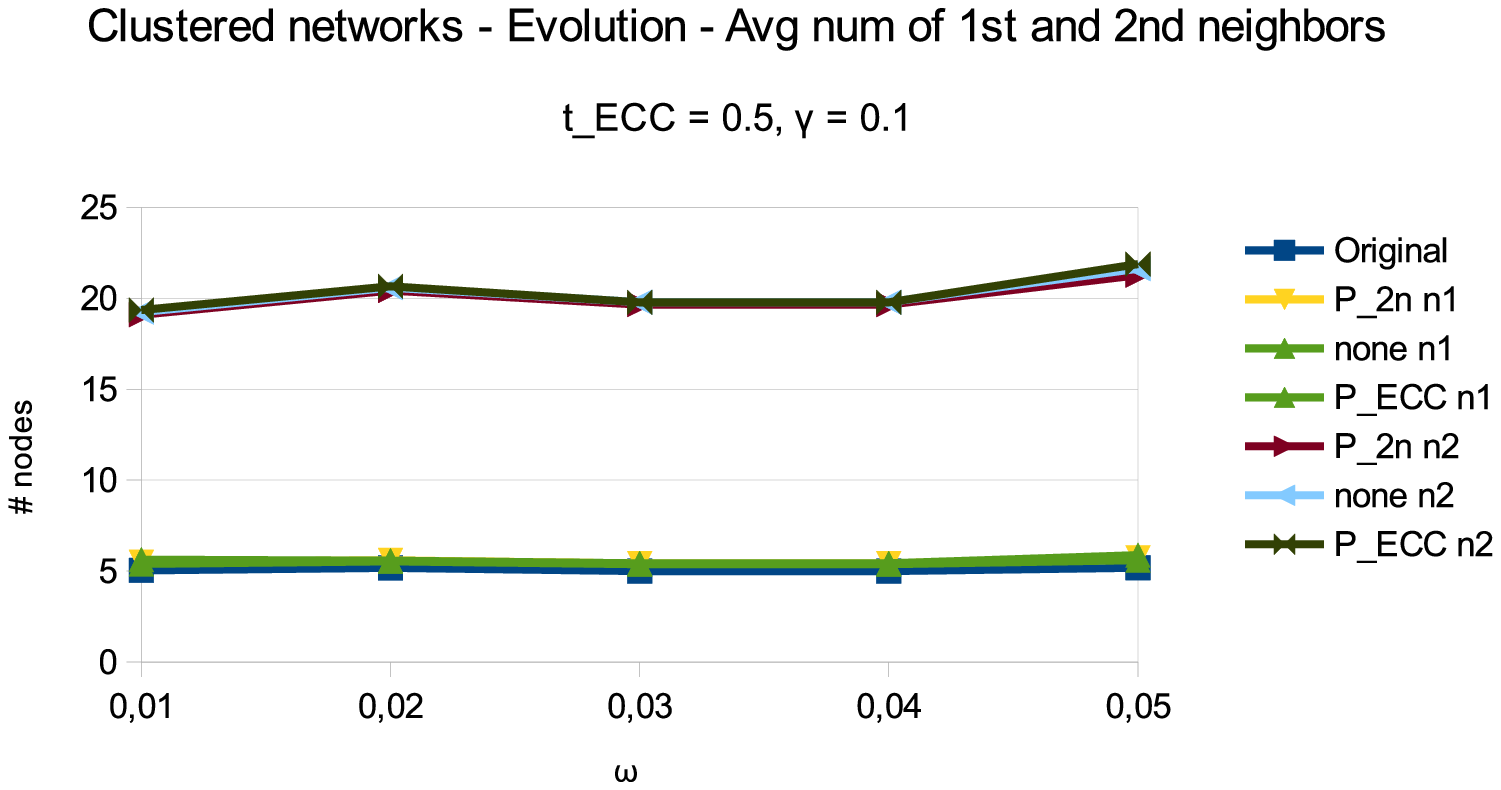}
   \caption{Clustered networks: average size of the main components, average amount of $1$st neighbors (referred as ``n1'') and $2$nd neighbors (referred as ``n2''), during the evolution of the network.}
   \label{fig:clus_evol}
\end{figure*}

\begin{figure*}[t]
   \centering
   \includegraphics[width=.48\linewidth]{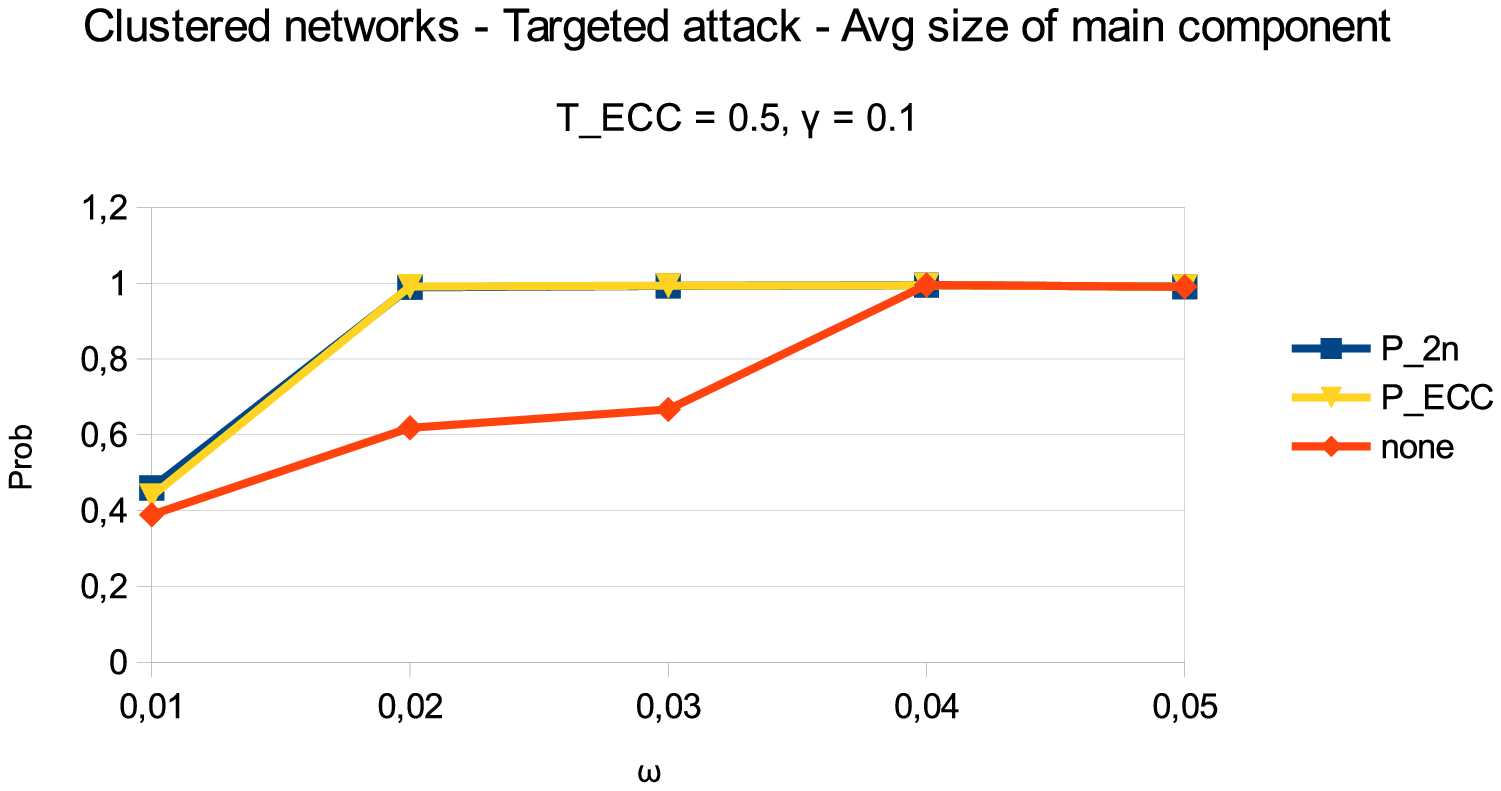}
   \includegraphics[width=.48\linewidth]{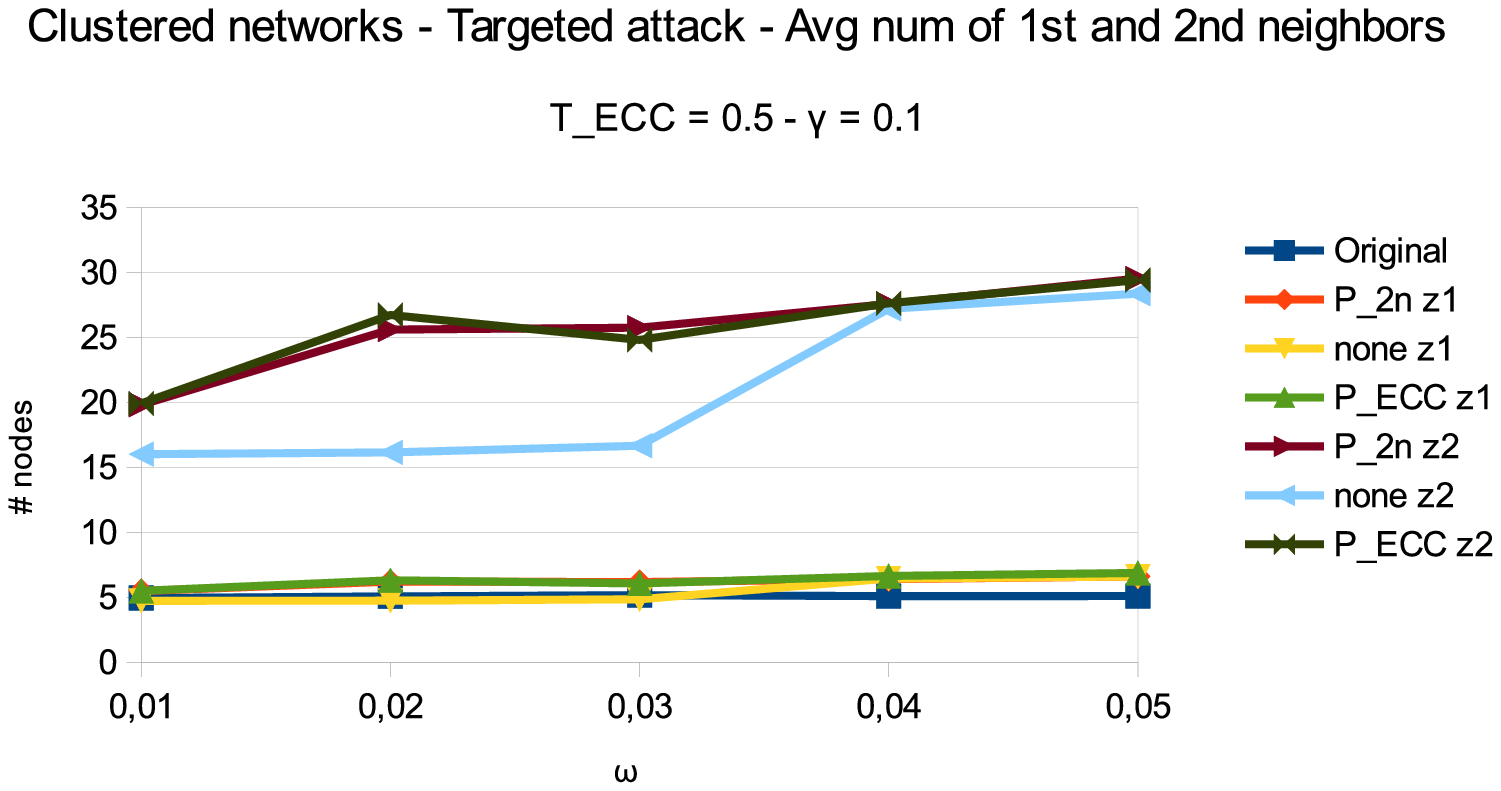}
   \caption{Clustered networks: average size of the main components, average amount of $1$st neighbors (referred as ``n1'') and $2$nd neighbors (referred as ``n2''), under a targeted attack.}
   \label{fig:clus_targ}
\end{figure*}

\begin{figure*}[t]
   \centering
   \includegraphics[width=.37\linewidth]{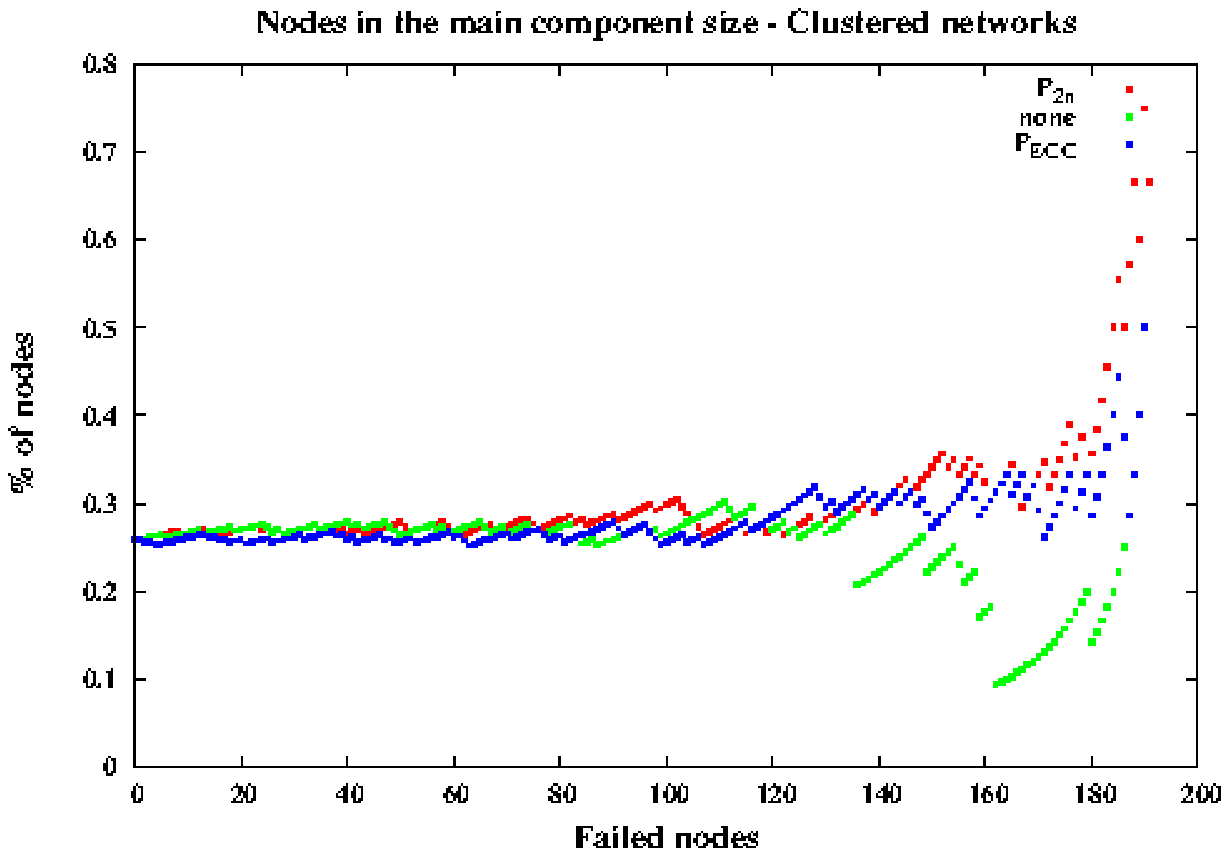}
         \hspace{1.5cm}
   \includegraphics[width=.37\linewidth]{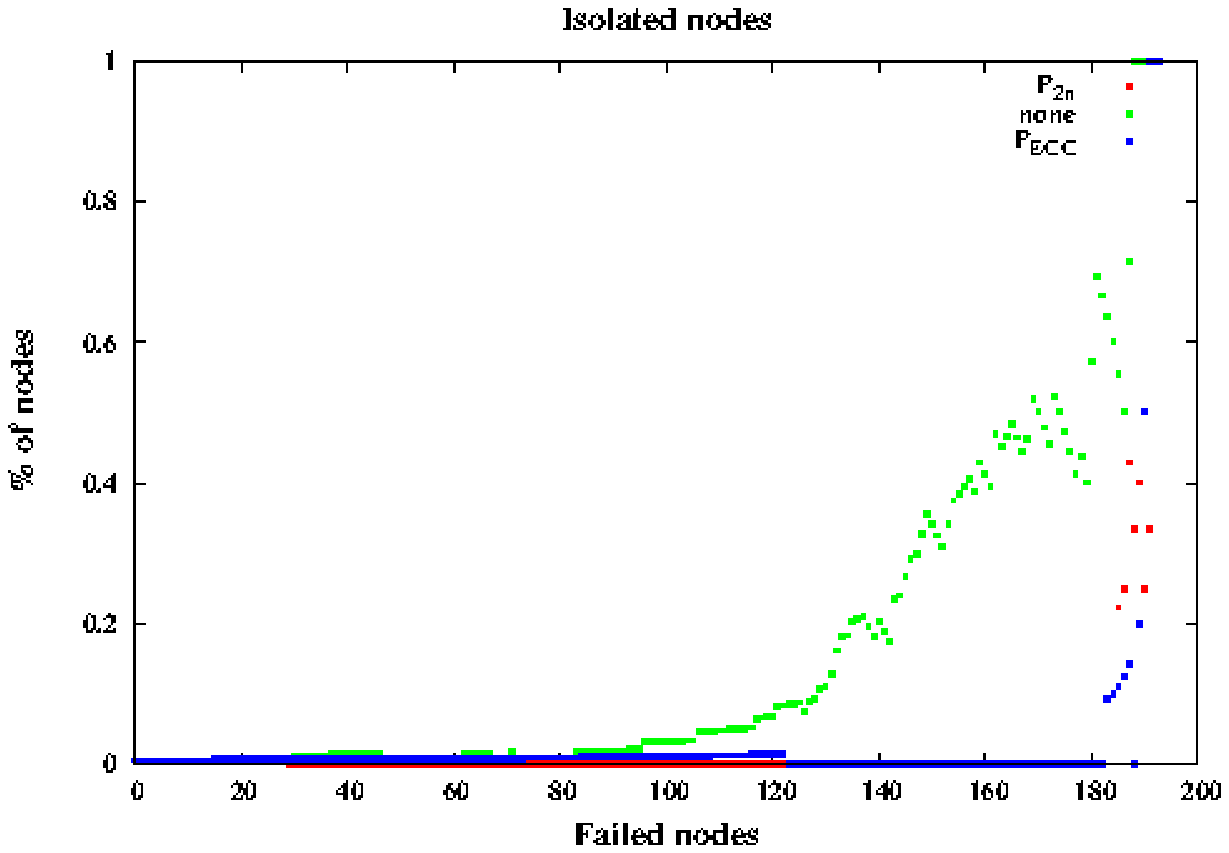}
   \caption{Clustered networks -- progressive node failures: Amount of nodes in the main component, isolated nodes.}
   \label{fig:clus_fail}
\end{figure*}

As concerns clustered networks, Figures \ref{fig:clus_evol}--\ref{fig:clus_fail} report the same metrics mentioned above for uniform nets. In this case, it is clear that a random removal of nodes (``evolution'' simulation mode) should not alter significantly the topology; hence, no particular benefits are evident from the use of $P_{2n}$ and $P_{ECC}$ w.r.t.~``none'' (see Figure \ref{fig:clus_evol}).
Indeed, the loss of contribution of the node that is selected to fail is in some sense balanced by the one that joins the net.
Things go differently when the targeted attack is employed (Figure \ref{fig:clus_targ}). Indeed, in this case the removal of a node that is particularly important for the interconnection among clusters easily leads to a disconnection of some components (left chart), and a reduction of neighbors (right chart) for the ``none'' protocol. Conversely $P_{2n}$ and $P_{ECC}$ behave quite well, preventing partitions in the network.

Figure \ref{fig:clus_fail} shows results obtained with the ``failures only'' simulations. As for uniform networks, the chart on the left shows the amount of nodes that remain in the main component during the evolution, while nodes continuously fail.
In this case, the network was disconnected, in the sense that the main component comprised only a percentage (slightly over $25\%$) of the whole set of nodes.
We might see that in this case, the main component size remains almost stable, for all the three protocols, until a half of the nodes become disconnected. This is due to the fact that the random choice of the failing nodes would privilege those nodes that were not in the bigger component (that includes less than the $30\%$ of nodes). However, in the last part of the simulation run the ``none'' protocol experiences a progressive decrement of nodes in the main component, since the main component is partitioned by the failures of its nodes. Conversely, the size of the main component increases for $P_{2n}$ and $P_{ECC}$. This is explained by the presence of the failure management protocols, that prevent the partition of the components.
The chart on the right of the figure confirms this, by reporting the amount of isolated nodes. While this amount progressively increases in the ``none'' protocol, in $P_{2n}$, $P_{ECC}$ the percentage of isolated nodes remains negligible for the main part of the simulation. Only at the end of the simulation some non-negligible amount of isolated nodes appears. This is explained by the fact that after a while some (minor) component remained composed of a single node (all the other nodes already failed).

\subsubsection{Scale-free Networks}
\begin{figure*}[t]
   \centering
   \includegraphics[width=.48\linewidth]{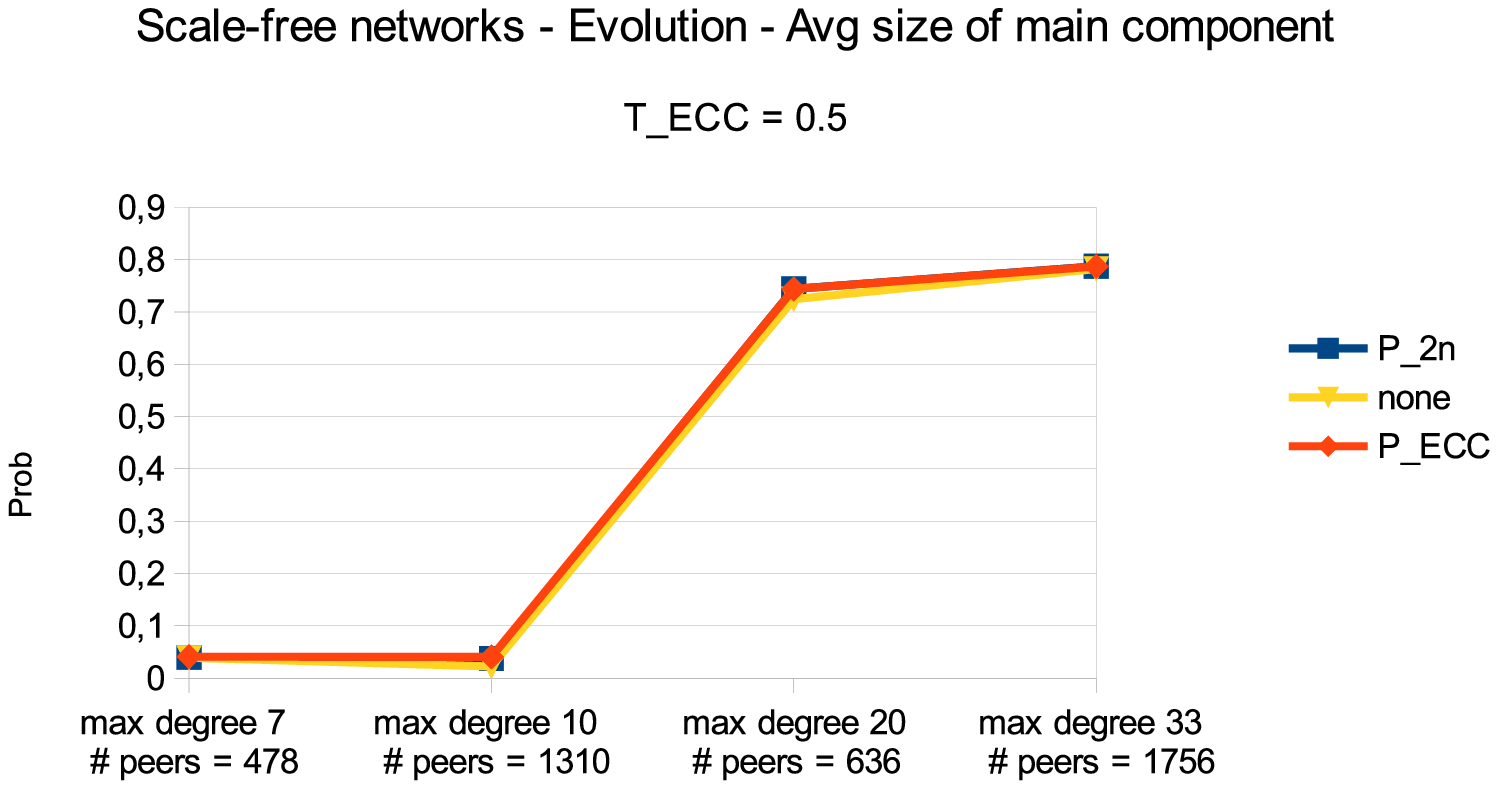}
   \includegraphics[width=.48\linewidth]{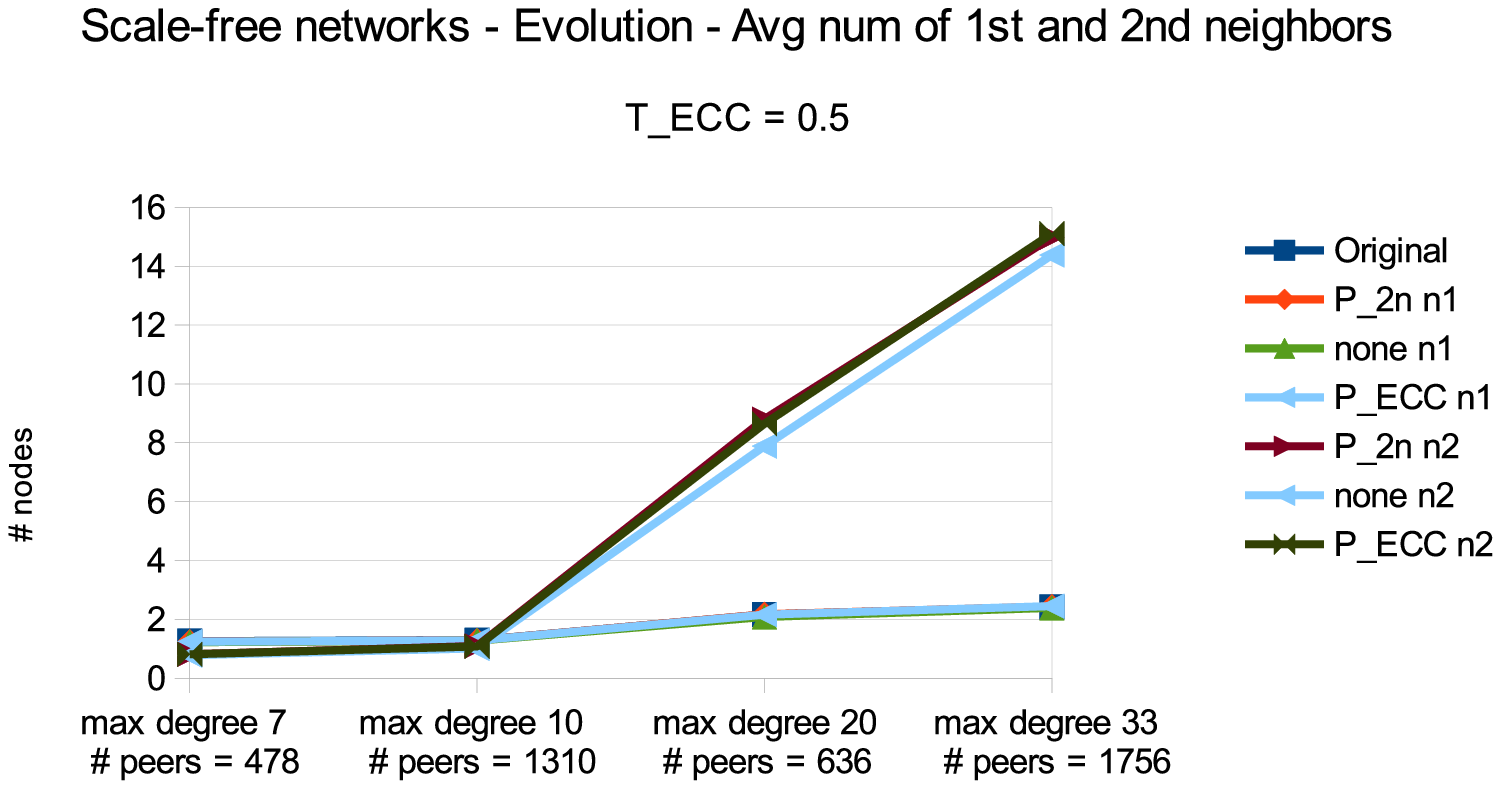}
   \caption{Scale-free networks: average size of the main components, average amount of $1$-neighbors (referred as ``n1'') and $2$-neighbors (referred as ``n2''), during the evolution of the network.}
   \label{fig:sf_evol}
\end{figure*}

\begin{figure*}[t]
   \centering
   \includegraphics[width=.48\linewidth]{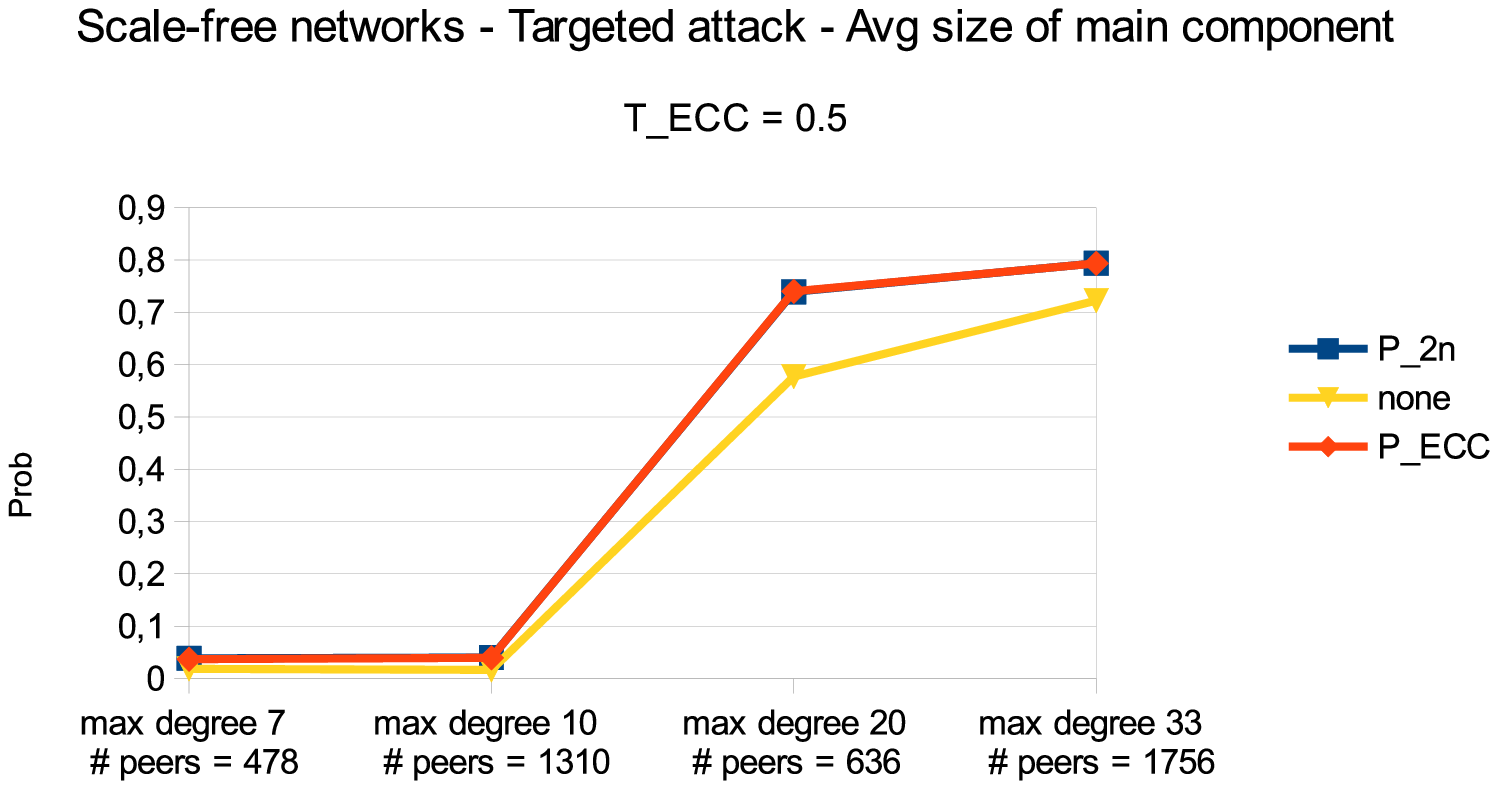}
   \includegraphics[width=.48\linewidth]{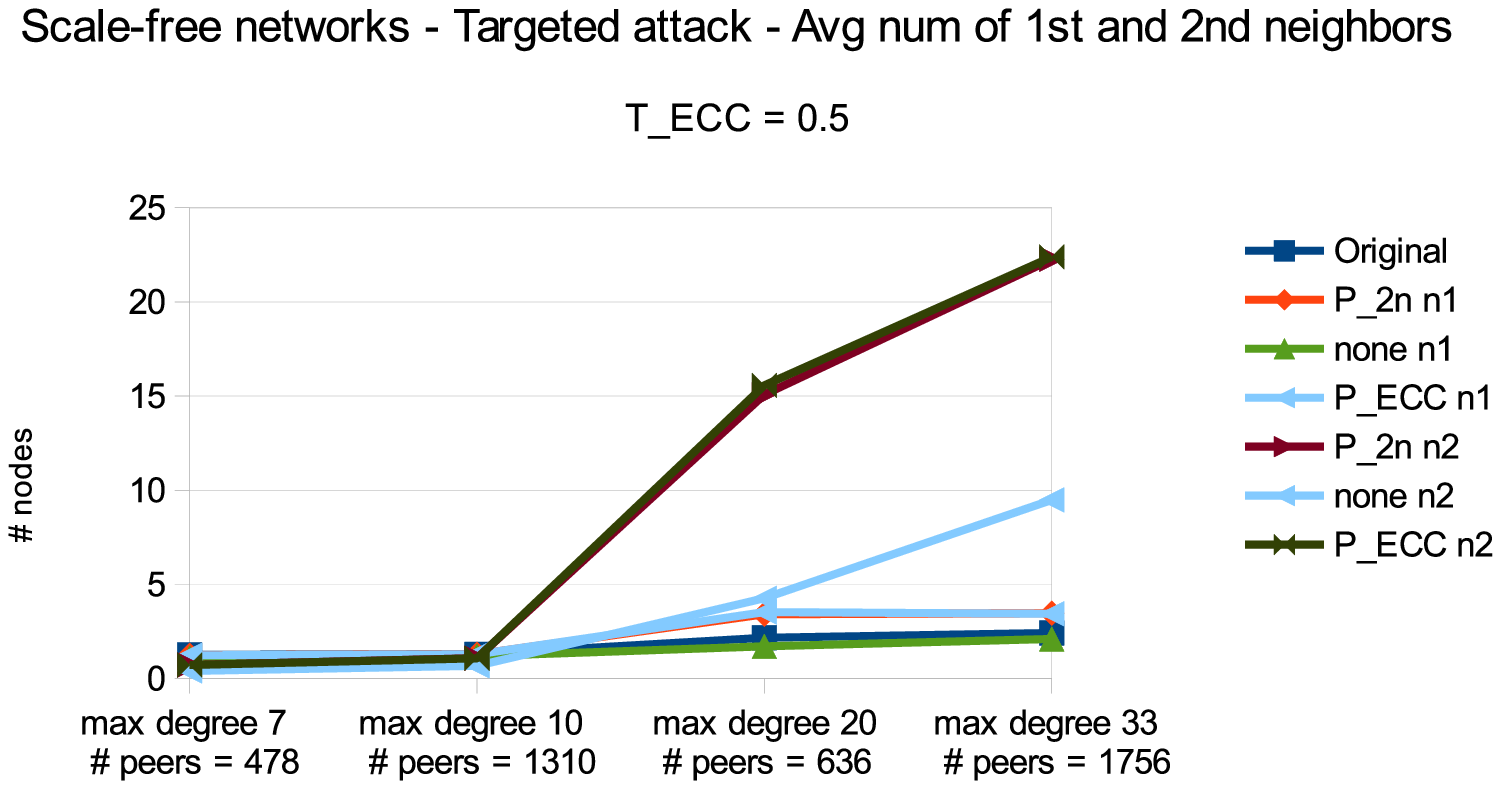}
   \caption{Scale-free networks: average size of the main components, average amount of $1$-neighbors (referred as ``n1'') and $2$-neighbors (referred as ``n2''), under a targeted attack.}
   \label{fig:sf_targ}
\end{figure*}

\begin{figure*}[t]
   \centering
   \includegraphics[width=.37\linewidth]{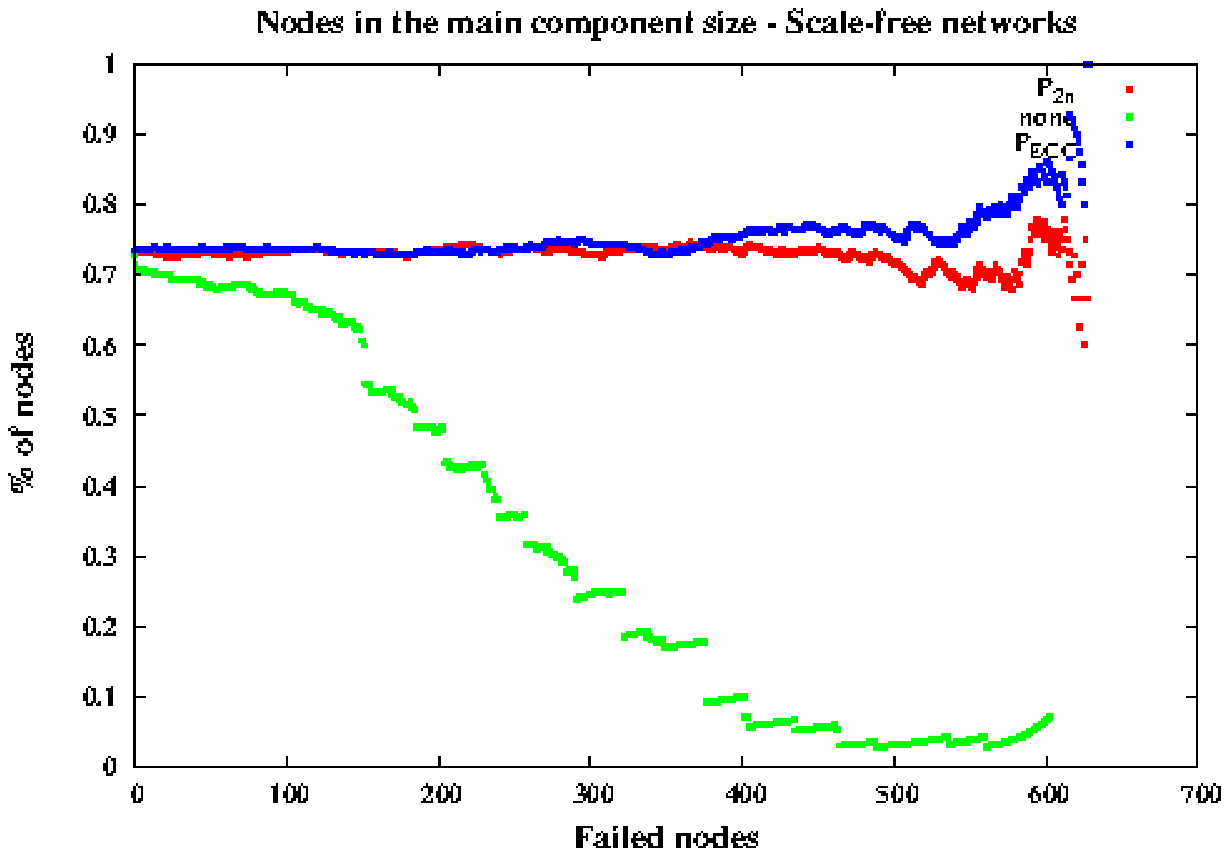}
      \hspace{1.5cm}
   \includegraphics[width=.37\linewidth]{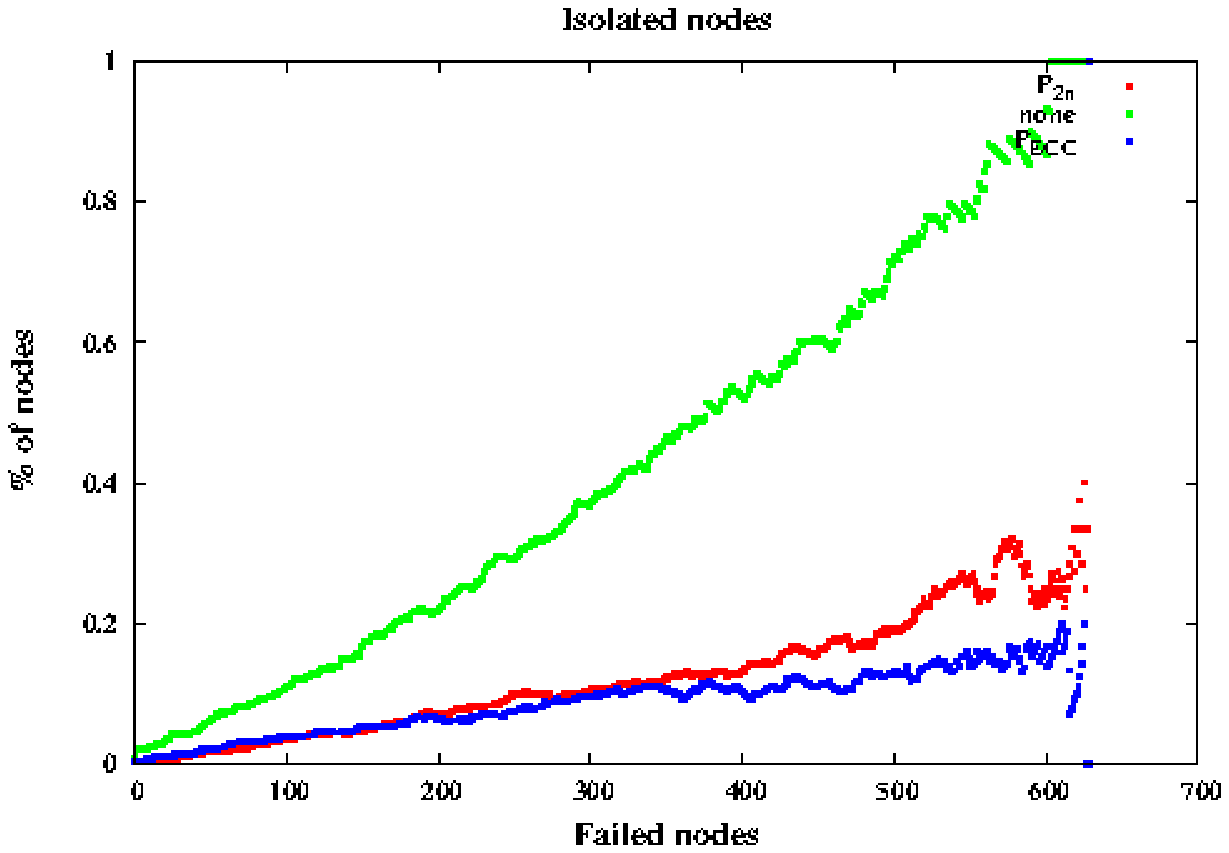}
   \caption{Scale-free networks -- progressive node failures: Amount of nodes in the main component, isolated nodes.}
   \label{fig:sf_fail}
\end{figure*}

Figures \ref{fig:sf_evol}--\ref{fig:sf_fail} show results for scale-free networks.
When the network evolves with random node faults (Figure \ref{fig:sf_evol}), it is more probable that nodes are selected with lower degrees (since these are more frequent). It is well known that scale-free nets are particularly resilient to random node faults, and these tests confirm it. 
In the charts, we consider four different types of scale-free networks, having different number of nodes and a different maximum degree which can be associated to nodes. This comes from the initial construction scheme employed to build the network, that uses two parameters, that combined together shape the scale-free networks \cite{Aiello00arandom}.
It is possible to see that, because of this employed construction method, the first two (types of) networks, reported on the $x$ axis, have a very small maximum degree, w.r.t.~the net size. (Put in other words, 
the node degree distributions follow a power-law behavior, typical of scale-free networks, but there are no real hubs in the nets.) Hence these two networks are composed of several small components with a very low number of $1$st and $2$nd neighbors. 
Results are different for the other two networks, that have higher maximum degrees. Indeed, in these cases the main components comprise a majority of the network (higher than the $70\%$, in one case, and around the $80\%$ in the other case, as shown in the left chart in Figure \ref{fig:sf_evol}).
Anyway, with the random node faults of the ``evolution'' simulation mode, the three protocols provide similar results.

Conversely, things change if we employ a targeted attack simulation mode. It is well known that scale-free networks are vulnerable to a targeted attack where the hubs are the nodes selected to fail. Indeed, as can be seen in Figure \ref{fig:sf_targ}, the ``none'' protocol has a substantial decrement on the size of the main component and the average amount of (especially $2$nd) neighbors.
Instead, $P_{2n}$ and $P_{ECC}$ react to the failure of hubs by pushing nodes to create novel links with their lost $2$nd neighbors. Thus, the proposed protocols do represent a solution to react to the targeted attack problem.

When we pass to the ``failures only'' simulation mode, we observe that the failure management protocols provide benefits to scale-free nets. Figure \ref{fig:sf_fail} reports results for a simulation run on a scale-free network composed of $636$ nodes, with a maximum degree of $20$ (for those interested in the specific construction method \cite{Aiello00arandom}, it employs two parameters that in this case were set to $a=6$, $b=2$). 
By looking at the chart on the right, it is possible to see that the simulation starts with a main component composed of more than the $70\%$ of the nodes. In the ``none'' mode, the component size progressively loses all its nodes, while in the $P_{2n}$ and $P_{ECC}$ protocols, the main component maintains its size (which actually increases in percentage, upon failure of nodes outside the main component). Actually, in this case $P_{ECC}$ outperforms $P_{2n}$. This is confirmed by the chart on the right in the figure, that reports the amount of isolated nodes.

\subsubsection{On the Network Topologies}
These results confirm that the two self-healing protocols allow to preserve network connectivity.
Now, an interesting question might be if these protocols alter the network topology, due to the creation of novel links among nodes. 
The obtained results show that in general the average degree (and its standard deviation) do not change significantly for uniform and clustered networks. As to scale-free networks, we did not noticed any particular alterations under the ``evolution'' simulation mode. 
However, we already observed that targeted attacks affect severely scale-free networks, since the failure of a hub cancels several links connecting different portions of the net. 

In this case, it is likely that the failure management would involve the creation of (many) novel links depending on the local needs of nodes, not necessarily following a preferential attachment strategy (or any other strategy typical of a scale-free net). To sum up, the failure management scheme could have some impact on a scale-free topology.

\begin{figure*}[t]
   \centering
   \subfigure[Original net]{\includegraphics[width=.37\linewidth]{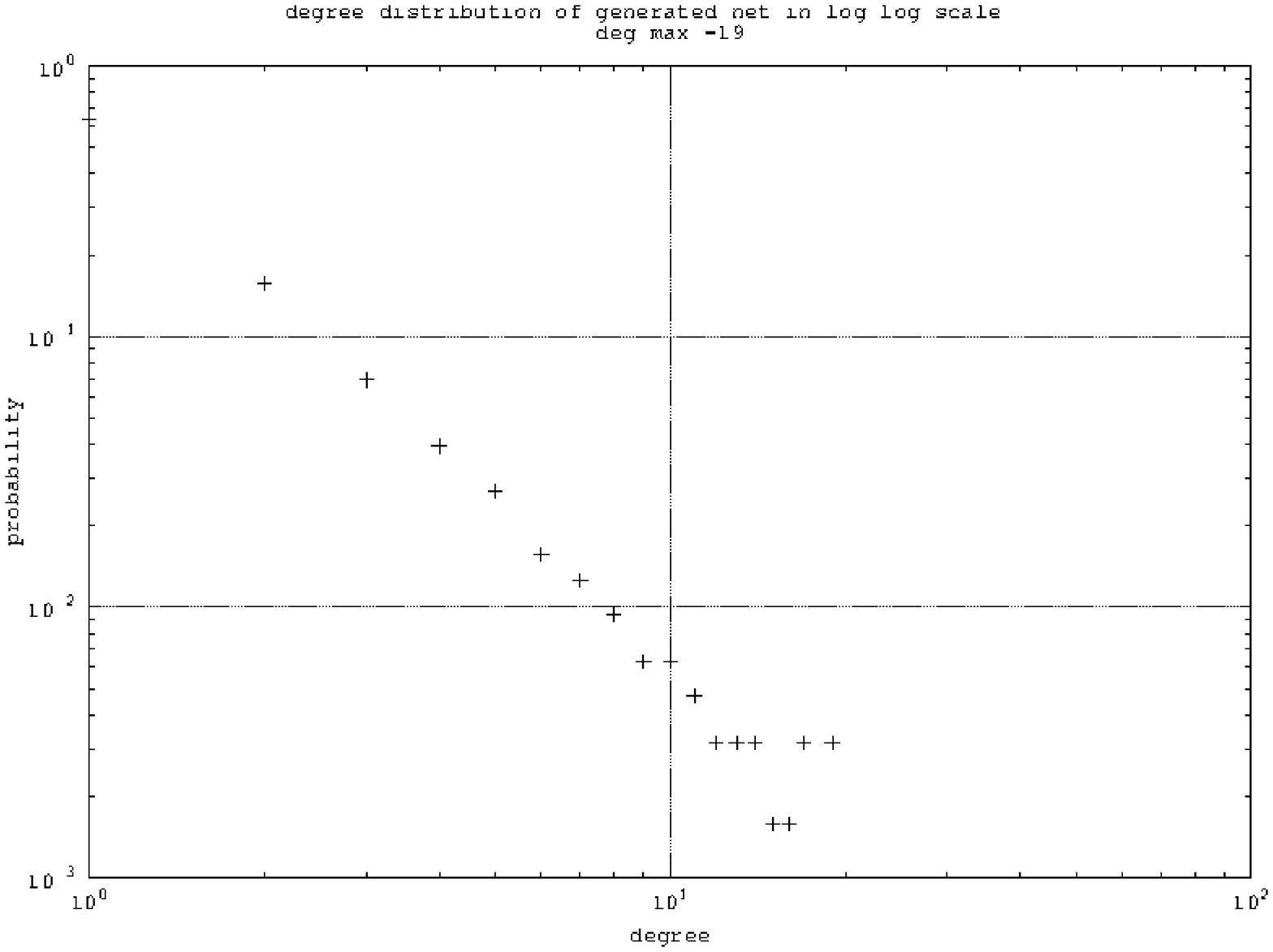}}
   \hspace{1.5cm}
      \subfigure[none]{\includegraphics[width=.37\linewidth]{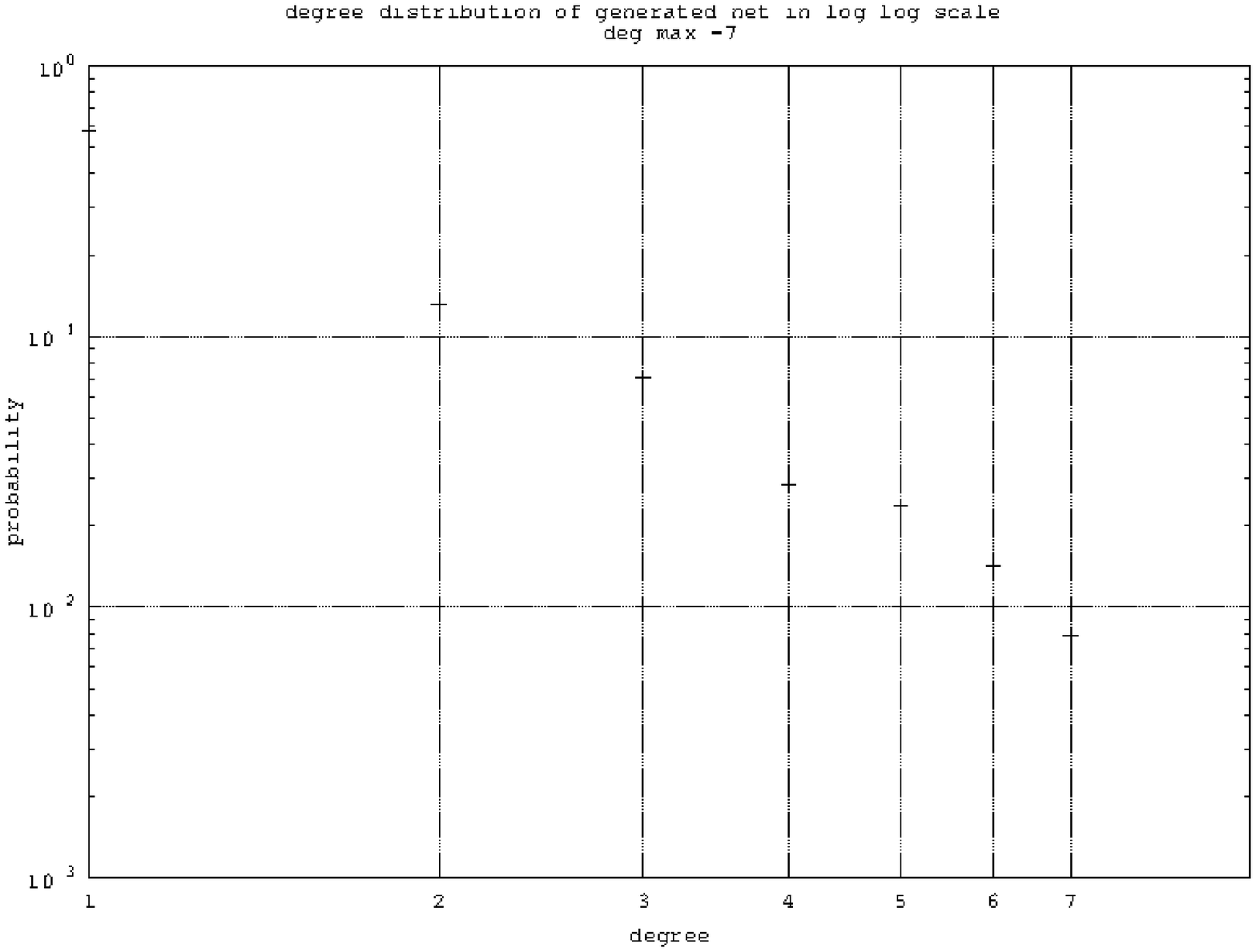}}
      \subfigure[$P_{2n}$]{\includegraphics[width=.37\linewidth]{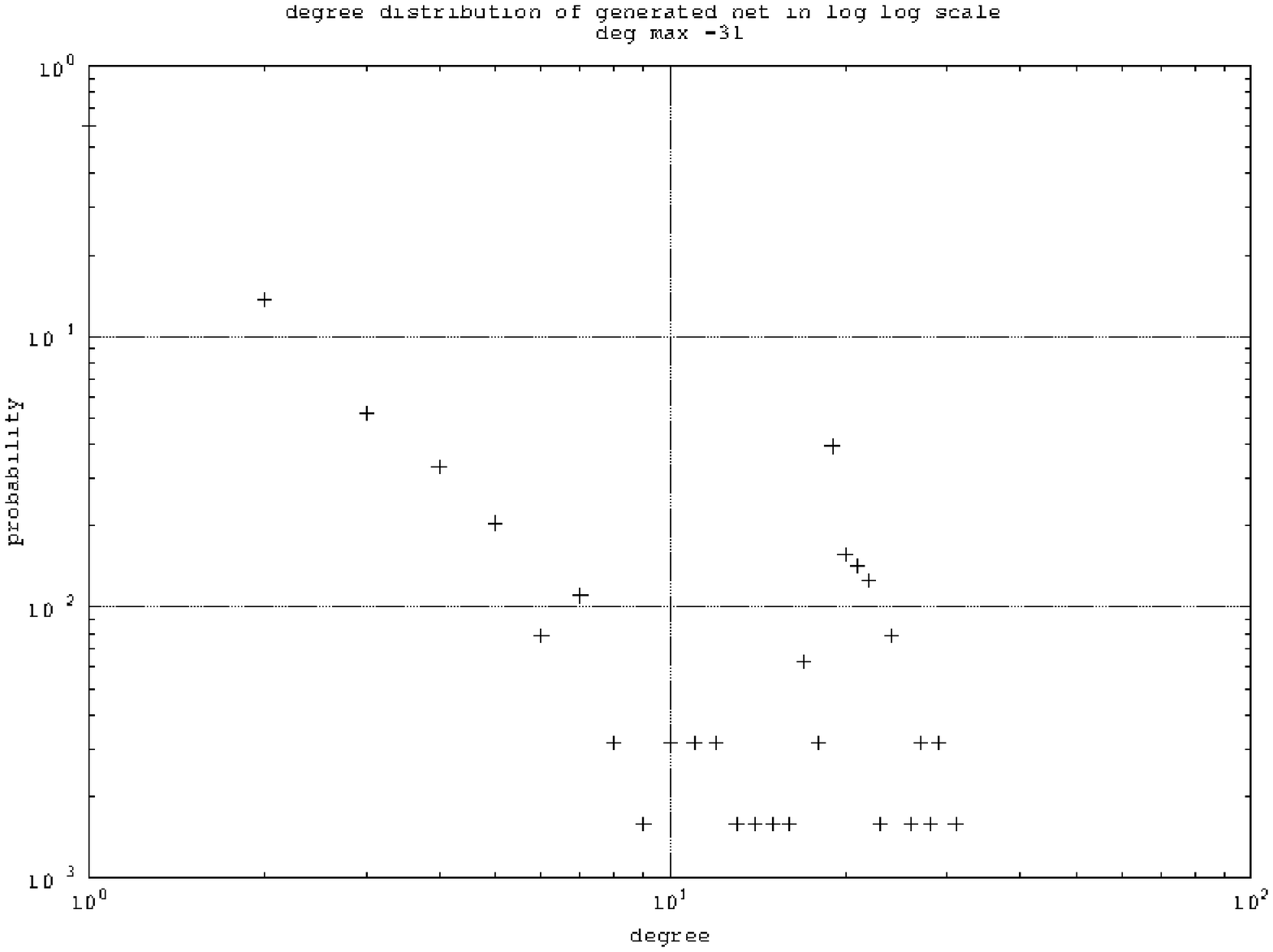}}
   \hspace{1.5cm}
      \subfigure[$P_{ECC}$]{\includegraphics[width=.37\linewidth]{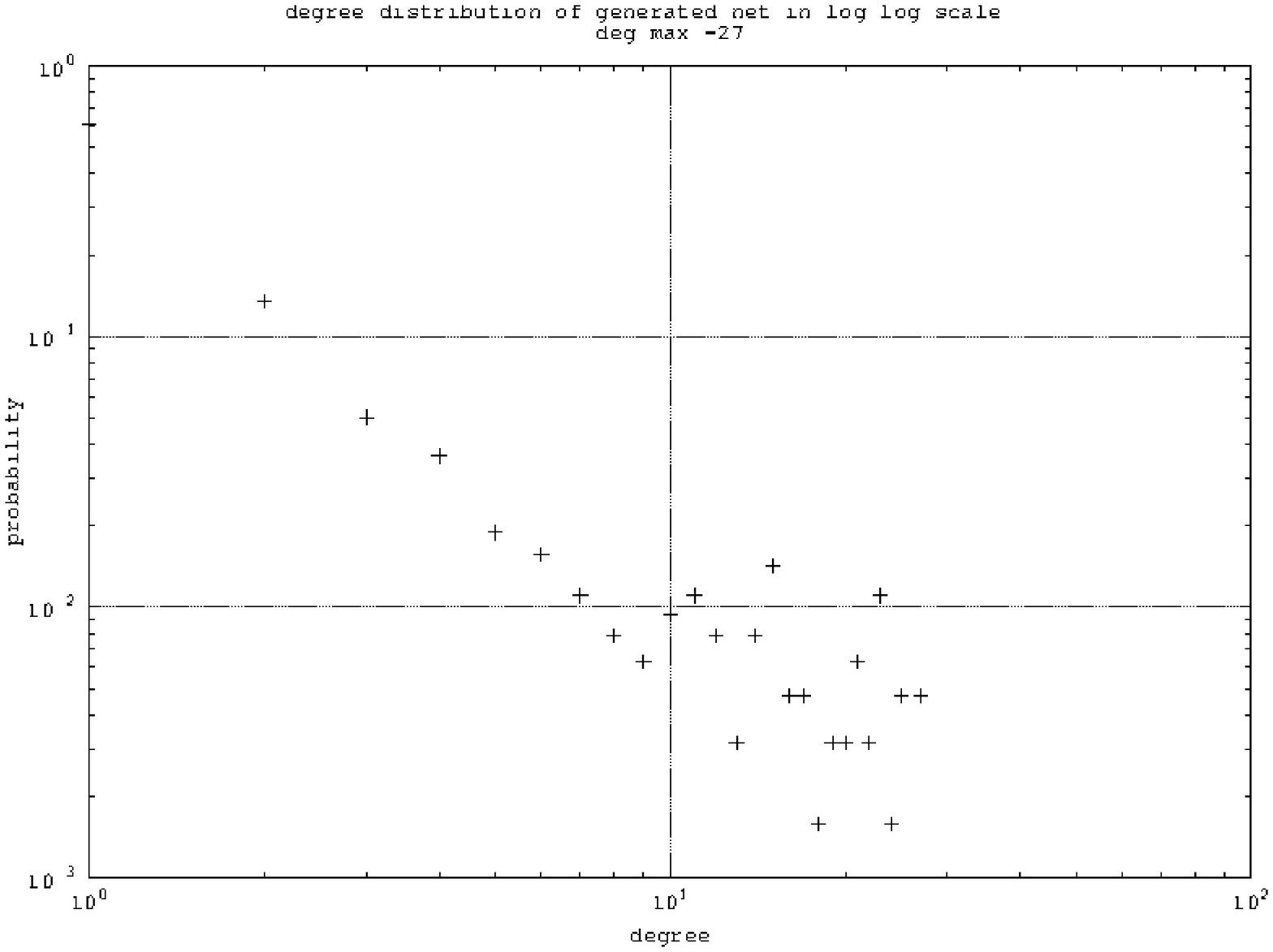}}
   \caption{Scale-free networks -- degree distribution under a targeted attack using the three approaches.}
   \label{fig:sf_deg_dist}
\end{figure*}

With this respect, it is interesting to look at Figure \ref{fig:sf_deg_dist}, that reports the degree distribution, for a given network in its original form (i.e.~before starting the simulation run), and the resulting distribution using the three compared mechanisms, when a targeted attack is simulated. We repeated the same procedure for several networks and different simulation runs, that always led to similar conclusions. In this case, the degree distributions were calculated taking a snapshot of the network after $50$ simulation steps.
In the Figure, each chart reports the degree distribution in a log-log scale, which allows to assess if the degree distribution follows a power law.
In fact, the original network follows a power law distribution (first chart in the figure). The network obtained with the ``none'' protocol maintains the power law distribution; however, the degree probabilities are reduced. Indeed, we already observed in the previous figures, that the network loses several of its properties, in terms of connectivity.

Not very surprisingly, $P_{2n}$ produces a degree distribution which is clearly different to a power law. In fact, when a hub fails, all its neighbors start the recovery procedure and create some links with their lost $2$nd neighbors. Put in other words, nodes create novel links in order to cope with the hub failure, sharing the load which was associated to it.
Depending on the type of P2P system and on the applications run on top of it, the fact that the final topology is altered may represent an acceptable or unacceptable side effect. However, it should be recalled that we are in the presence of a perpetuated targeted attack that, without the use of a self-healing protocol, would affect severely the scale-free network, since it prevents the creation of novel hubs.

As concerns, $P_{ECC}$, things are different. In fact, the obtained degree distribution is not perfectly linear; there are some higher degrees with higher probabilities than expected. However, a linear trend is more evident w.r.t.~$P_{2n}$. The idea is that in this case, using the notion of $ECC$ and the reconfiguration phase that eliminates unneeded links (when multiple triangles are created that did not existed before), a node has an incentive to maintain an important connection that has been lost.
This means that $P_{ECC}$ provides the same results in terms of reliability and resilience w.r.t.~$P_{2n}$, with a reduced impact on the network topology.

\section{Conclusions}\label{sec:conc}

This paper focused on two distributed mechanisms that can be executed locally by peers in an unstructured overlay, in order to cope with node failures and augment the resilience of the network.
The two self-healing protocols require knowledge of $1$st and $2$nd neighbors. 
Outcomes confirm that it is possible to augment resilience and avoid disconnections in unstructured P2P overlay networks. 

In particular, while both schemes are of help to avoid network disconnections, our results suggest that the use of the Edge Clustering Coefficient (ECC) has the advantage of reducing possible alterations on the network topologies, which may arise during the self-healing phase.
In fact, ECC provides an idea of how much inter-communitarian a link is. It can be thus exploited to: i) replace lost important links with novel ones after some failures, ii) (if needed) remove those (novel) links that might augment excessively the degree of some node and the amount of triangles it belongs. 


%
%
\bibliographystyle{abbrv}


\end{document}